\documentclass[aps,pre,a4paper,preprint,amsmath,showpacs]{revtex4-1}
\usepackage{graphicx,amssymb}
\usepackage[show]{ed}
\usepackage{xcolor}
\usepackage[latin1]{inputenc}
\usepackage{color,soul}

\begin{document}
\title{Reaction-diffusion and reaction-subdiffusion equations on arbitrarily evolving domains}
\author{E. Abad}
\email{eabad@unex.es}
\affiliation{Departamento de F\'{i}sica Aplicada and Instituto de Computaci\'on Cient\'{i}fica Avanzada, Centro Universitario de M\'erida, Universidad de Extremadura, 06800 M\'erida, Spain}

\author{C. N. Angstmann}
\email{c.angstmann@unsw.edu.au}
\affiliation{School of Mathematics and Statistics, UNSW, Sydney NSW, 2052, Australia}

\author{B. I. Henry}
\email{b.henry@unsw.edu.au}
\affiliation{School of Mathematics and Statistics, UNSW, Sydney NSW, 2052, Australia}

\author{F. {Le Vot}}
\email{felipelevot@unex.es}
\affiliation{Departamento de F\'{i}sica and Instituto de Computaci\'on Cient\'{i}fica Avanzada, Universidad de Extremadura, 06071 Badajoz, Spain}

\author{A. V. McGann}
\email{a.mcgann@unsw.edu.au}
\affiliation{School of Mathematics and Statistics, UNSW, Sydney NSW, 2052, Australia}

\author{S. B. Yuste}
\email{santos@unex.es}
\affiliation{Departamento de F\'{i}sica and Instituto de Computaci\'on Cient\'{i}fica Avanzada, Universidad de Extremadura, 06071 Badajoz, Spain}

\date{\today}

\begin{abstract}
Reaction-diffusion equations are widely used as the governing evolution equations for modeling many physical, chemical, and biological processes. Here we derive reaction-diffusion equations to model transport with reactions on a one-dimensional domain that is evolving. The model equations, which have been derived from generalized continuous time random walks, can incorporate complexities such as subdiffusive transport and inhomogeneous domain stretching and shrinking. A method for constructing analytic expressions for short time moments of the position of the particles is developed and moments calculated from this approach are shown to compare favourably with results from random walk simulations and numerical integration of the reaction transport equation. The results show the important role played by the initial condition. In particular, it strongly affects the time dependence of the moments in the short time regime by introducing additional drift and diffusion terms. We also discuss how our reaction transport equation could be applied to study the spreading of a population on an evolving interface.
\end{abstract}
\maketitle

\section{Introduction}
Reaction-diffusion partial differential equations have been widely employed to provide mathematical models across many physical, chemical and biological processes \cite{O1980,B1986,M2001}, with classic applications including the spread of bushfires, the development of animal coat patterns, and the spread of epidemics. In recent decades the fundamental development of reaction-diffusion equations has focussed on extensions to incorporate physical complexities in two key areas;  anomalous subdiffusion \cite{HW2000,HLW2006,SSS2006,LHW2008,F2010,AYL2010,YAL2010,ADH2013mmnp},
and domain growth
\cite{CGK1999,M2001,CM2001,BYE2010,WBGM2011,YBEM2012,SSMB2015,YAE2016,MB2018,VEAY2018,GKK2019}.
Anomalous subdiffusion, which has been reported in numerous experimental observations \cite{BWM2000,LB2003,WEKN2004,SWDA2006,MW2010,AYYHY2011,EK2011,S2012,JTHDB2016}, refers to diffusion processes in which the mean square displacement scales as a sublinear power law in time.
Domain growth is used generically to refer to stretching and shrinking of the domain over time.
In this work we have developed reaction-diffusion equations to allow for the possibility of including both features, subdiffusion and domain growth, simultaneously.

The study of subdiffusion with reactions on growing domains is still in its infancy, but some important steps have already been made. In this context, the main tool used so far are continuous time random walk (CTRW) models.  Such models can be generalized at various levels. One possibility is the inclusion of chemical reactions, but this should be done carefully, since the correct form of the evolution equations is not always the most intuitive one \cite{HW2000,HLW2006,SSS2006,LHW2008,F2010,AYL2010,YAL2010,ADH2013mmnp}.

We now proceed to give a brief overview of recent progress and how it relates to the goals of this paper.  Le Vot, Abad and Yuste  \cite{VAY2017} used a CTRW approach to obtain evolution equations for unbiased diffusion, including anomalous diffusion, in uniformly expanding or contracting media. Later, these results were extended by Le Vot and Yuste \cite{VY2018} to account for the effect of a biasing force field. Another significant advance was made by Angstmann, Henry and MacGann \cite{AHM2017pre}, which employed a generalized CTRW formalism  \cite{MW1965,MK2000} to deal with the general case of random walkers that move diffusively or subdiffusively in domains with inhomogeneous growth/contraction rates. In particular, this means that some regions of the domain may expand, while others may simultaneously shrink.  The next step was the inclusion of chemical reactions in models with normal diffusion. This was done in Refs.~\cite{VEAY2018, EYAV2018, AEVY2019}, where evolution equations for encounter-controlled reactions between reactants performing normal diffusive walks in uniformly evolving domains were derived and solved for the special cases of particle coalescence and annihilation. Here, we take the formalism developed in Ref.~\cite{AHM2017pre} as a starting point by including chemical reactions in the transport equations that hold for arbitrarily evolving domains. Such reactions are not restricted to death processes as in \cite{VEAY2018, EYAV2018, AEVY2019}, but may also involve particle birth.

We begin in Section II by setting up a formalism and  co-ordinate system to describe non-uniform domain growth.
In Section III we first consider a single particle CTRW on a growing domain, with a death  probability between steps.
The CTRW on the growing domain is mapped to an auxiliary CTRW in terms of comoving coordinates, i.e., spatial coordinates referring to the initial fixed domain. Then, the master equation is obtained for this auxiliary process. We subsequently consider an ensemble of particles undergoing CTRWs with death probabilities and creation probabilities between steps and the master equation for the auxiliary CTRW on the fixed domain with reactions is obtained. In Section IV we derive the diffusion limits of the auxiliary CTRW master equation with a jump length density corresponding to unbiased jumps of fixed length on the evolving domain, and with two distinct waiting time densities,  exponential and Mittag-Leffler. These densities are known to limit to standard diffusion, and subdiffusion respectively on fixed domains. The fixed length jump on the growing domain maps to a space- and time- varying jump length for the auxiliary CTRW on the fixed domain. An iterative method for evaluating moments to higher orders in time is introduced and moments are evaluated for special cases. These moments are first evaluated for the auxiliary process and then mapped to moments on the evolving domain. The moment calculations are shown to compare favourably with numerical simulations, both on the auxiliary domain and on the evolving domain. In Section V we map the diffusion limit equations for the auxiliary CTRW on the fixed domain back to the evolving domain. This yields reaction-diffusion equations for systems undergoing standard diffusion, or subdiffusion, including reactions, on arbitrarily evolving domains.
We present a physical example in Section VI and we conclude with a brief summary and outlook in Section VII.

\section{Spatial and Temporal Domain Evolution Function}
To obtain the governing evolution equations for diffusing particles on an evolving domain it is useful to establish a mapping between points on the evolving domain at time $t$ and points on the initial fixed domain at time $t=0$  \cite{AHM2017pre}.

For simplicity, in the following, we will assume a finite domain of initial size $L(t=0)\equiv L_0$. At a given time $t$ we associate each point on the evolving domain $y(t)\in[0,L(t)]$, with a point on the initial domain $x\in[0,L_0]$ through a one-to-one mapping $y(t)=\bar g(x,t)$. It should be noted that, although in the present discussion we assume that the domain remains finite at all times, there is no problem in considering it as large as necessary, or even infinite.
Here, and subsequently, we use a bar to denote any function of the space variable $x$ on the original domain.
 Especially in cosmology, this $x$-co-ordinate is also termed ``comoving distance'' \cite{YAE2016,EYAV2018}.
The mapping must satisfy the initial condition $\bar g(x,0)=x$, the boundary conditions $\bar g(0,t)=0$ and $\bar g(L,t)=L(t)$, and the non-negativity condition $\bar g(x,t)\ge 0$.

Assuming that the domain is a differentiable manifold, the mapping $\bar g(x,t)$ can be represented uniquely in terms of a local growth rate function $\bar\mu(x,t)$. We partition the initial domain into $n$ intervals of equal length, $\delta x=L_0/n$. On the evolving domain, the length of each partitioned interval may change in time.
We let $\delta y_i (t)$ denote the length of the $i^{\mathrm{th}}$ partition at time $t$.
We can now define a local growth function $\bar\mu(x_i,t)$ through the  evolution equation \cite{AHM2017pre}
\begin{equation}
\frac{d\delta y_i}{dt}=\bar\mu(x_i,t)\delta y_i.
\label{Eq1}
\end{equation}

\begin{figure}
\includegraphics[width=0.7\textwidth]{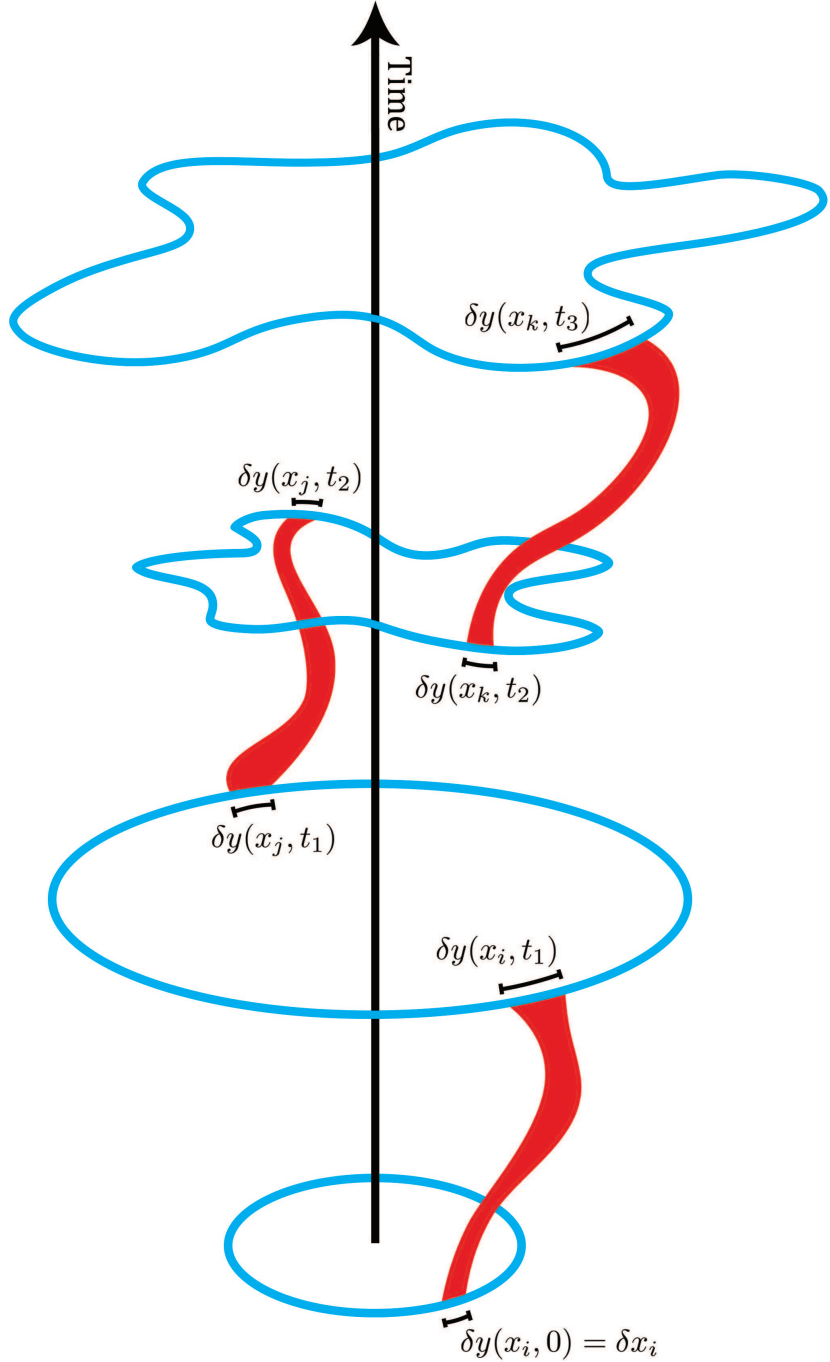}
\caption{Schematic illustration of an evolving domain. Here the domain initially expands uniformly as an ellipse before local expansions and contractions lead to a more irregular profile. The red shading is used to illustrate how particular intervals evolve in time with different local space- and time- dependent growth rates. Note that, although we have depicted the $1d$ domain as a ring, in general, the medium would be a line with two independent ends.}
\label{fig:domain}
\end{figure}

Integrating Eq.~\eqref{Eq1}, and using the initial condition $\delta y_i(0)=\delta x$ results in the expression,
\begin{equation}
\label{deltay}
\delta y_i(t)=\left[\exp \int_0^t \bar\mu(x_i,s)ds\right] \delta x= \bar \nu(x_i,t)\, \delta x,
\end{equation}
where, in the rightmost equation, the quantity
\begin{equation}
\bar \nu(x,t)\equiv e^{\int_{0}^{t}\bar\mu(x,s)ds}
\end{equation}
has been introduced. In the language of cosmology, this quantity is referred to as ``the scale factor''.
Summing all the $\delta y_i$ intervals and taking the limit as $n\rightarrow\infty$, produces the mapping between the domains, which can be expressed as a function of the local growth rate.
Explicitly,
\begin{equation}\label{ymux}
y=\lim_{n\rightarrow\infty} \sum_{i=0}^n \delta y_i=\int_0^x \bar \nu(z,t) dz=\bar g(x,t).
\end{equation}
The local growth rate function might be positive or negative, at different locations, allowing for local expansion and contraction respectively. This may result in a significant distortion of the original morphology of the domain, as illustrated schematically in Fig.~\ref{fig:domain} for the particular case of periodic boundary conditions. The evolution equations that we derive are unaffected by the complexity of the distortions, with the only restriction being that the domain remains a differentiable manifold.

Our derivation of the governing evolution equations for reaction and diffusion on an evolving domain utilizes the mapping between the growing domain and the fixed domain. Note that if $f(y,t)$ represents a function of the space variable on the evolving domain $y$, and $\bar f(x,t)$ represents the corresponding function on the fixed domain $x$, then these functions are related through the mapping $\bar g(x,t)$ via
\begin{equation}\label{eq_grow_transform}
f(y,t)=\bar{f}(\bar g^{-1}(y,t),t)=\bar{f}(x,t).
\end{equation}

The advantage in using the representation on the fixed domain is that this enables us to employ standard time derivatives, and then map them to the evolving domain.

It should be noted that our results are also valid for $d$-dimensional evolving domains when the diffusion processes along each Cartesian direction are independent, that is, when
\begin{equation}
\frac{d\delta y_i^{(j)}}{dt}=\bar\mu^{(j)}(x_i,t)\delta y_i^{(j)},
\end{equation}
where the superscript $j$ denotes the $j$-th Cartesian direction. An important special case is obtained when $\bar\mu^{(j)}(x_i,t)=\bar\mu(t)$, which corresponds to an isotropic $d$-dimensional exponential domain evolution.

\section{Master Equations for Auxiliary CTRWs}
\label{sec_der}
\subsection{Single Particles with Death Probabilities}
We begin by deriving the governing equation for  single particle diffusion on a growing domain with an associated death probability. Subsequently we consider  an ensemble of such particles and we also include birth events.
We let $q_n(y,t|y_0,0)$ denote the  probability per unit time for a particle that started at $y_0$ at time $t=0$ to arrive at $y$ at time $t$ after $n$ jumps, and we suppose that initially
\begin{equation}
q_0(y,t|y_0,0)=\delta(y-y_0)\delta(t-0^+).
\end{equation}
After $n+1$ steps, the arrival probability rate can be expressed with a recursion relation as,
\begin{equation} \label{eq_nsteps}
q_{n+1}(y,t|y_0,0)=\int_0^{L(t)}\int_0^t \Psi(y,y'',t,t')\theta(y',t,t')q_{n}(y',t'|y_0,0)  dt' dy'.
\end{equation}
The interpretation of this is that the probability rate for a particle to arrive at $y$ at time $t$ after $n+1$ steps is the probability rate for a particle that arrived at
$y'$ at an earlier time $t'$, then waited for a time $t-t'$, and survived death, before jumping to $y$ at time $t$.
The position $y''$ represents the location on the growing domain of a point at time $t$ that was at a point $y'$ at the earlier time $t'$. The term $\theta(y',t,t')$ is the probability of surviving death at a point on the domain, referenced as $y'$ at time $t'$, and $\Psi(y,y'',t,t')$ is the probability density for waiting a time $t-t'$ and transitioning from $y''$  to $y$ at time $t$.

The probability per unit time for a particle to arrive at $y$ at time $t$ after any number of steps  is obtained by summing over all $n$ in Eq.~\eqref{eq_nsteps}. This results in
\begin{equation}\label{eq_fluxq}
q(y,t|y_0,0)=q_0(y,t|y_0,0) +\int_0^{L(t)}\int_0^t \Psi(y,y'',t,t')\theta(y',t,t')q(y',t'|y_0,0)dt' dy',
\end{equation}
where
\begin{equation}
q(y,t|y_0,0)=\sum_{n=0}^\infty q_n(y,t|y_0,0).
\end{equation}
In Eq.~\eqref{eq_fluxq}, the quantity
\begin{equation}
\theta(y',t,t')=e^{-\int_{t'}^t \omega(y',s)ds}
\end{equation}
stands for the survival probability of a particle arriving at location $y'$ at time $t'$, where $\omega(y,s)\delta s$ is the (infinitesimal) probability of a particle dying between times $s$ and $s+\delta s$. At this stage, it is worth noting that, while the above mortality law may not be the most general one, it does not preclude a dependence of the form $\omega=\omega(\rho(y,t))$, i.e., an explicit dependence of the survival probability $\theta$ on the probability density (or  ``concentration'') $\rho(y,t)$ of finding a particle at position $y$ at time $t$.

We assume that the transition probability density is composed of two independent probability densities:
the jump density, $\lambda(y,y'')=\lambda(y-y'')$,  for a jump of length $y-y''$, and the waiting time density, $\psi(t,t')=\psi(t-t')$, for a particle to wait for $t-t'$ time before jumping. Hence we write,
\begin{equation}
\Psi(y,y'',t,t')=\lambda(y-y'')\psi(t-t').
\end{equation}

The probability of finding a particle, undergoing a CTRW on a growing domain, in the infinitesimal volume interval $[y'',y''+dy'']$ at time $t$ can be written as
\begin{equation} \label{eq_rho0}
\begin{split}
\rho(y'',t|y_0,0) dy''=\int_{0}^{t}\Phi(t-t')\theta(y',t,t')q(y',t'|y_0,0) \, dy' dt',
\end{split}
\end{equation}
where
\begin{equation}
\Phi(t-t')=1-\int_0^{t-t'}\psi(s)ds,
\end{equation}
is the jump probability survival function, and $\rho(y'',t|y_0,0)$ is the probability density for being at $y''$ at time $t$.
In order to formulate a master equation for the evolution of $\rho(y,t|y_0,0)$, we find it convenient to consider an auxiliary CTRW process on the initial fixed domain \cite{AHM2017pre}. One advantage of this is that standard time derivatives can be carried out on the fixed domain and the corresponding functions can then be mapped back
 to the growing domain. A second advantage is that the equations for the auxiliary process on the fixed domain can be solved in this co-ordinate system and the solutions can then be mapped on to the evolving domain.
First we relate the densities on the evolving domain to associated densities on a fixed domain using the transformation of Eq.~\eqref{ymux}. Explicitly, the probability of finding a particle at time $t$ within the interval $[y,y+dy]$ is given by
\begin{eqnarray} \label{eq_rhomap}
\rho(y,t|x_0,0)\, dy&=&\rho(y,t|x_0,0)\, \frac{dy}{dx}\, dx\nonumber\\
&=&\rho(g(x,t),t|x_0,0) \bar \nu(x,t) \,dx\nonumber\\
&=&\bar\rho(x,t|x_0,0) \bar \nu(x,t)\, dx \nonumber \\
&=&\rho(x,t|x_0,0) \, dx.
\end{eqnarray}
The quantity $\rho(x,t|x_0,0) dx$ defined in the last line has a clear
physical interpretation: it is the probability of finding a particle at time $t$ within the interval $[x,x+dx]$
on the fixed domain, $[x,x+dx]$ being the $x$-interval corresponding to $[y,y+dy]$ at time $t$.
Because of probability conservation, one has
the relation $\rho(y,t|x_0,0)=(dy/dx)\rho(x,t|x_0,0)$.
Similarly, one has
\begin{equation}
q(y,t|x_0,0)\, dy=q(x,t|x_0,0) \, dx =\bar q(x,t|x_0,0) \bar \nu(x,t) \, dx.
\end{equation}
Along the same lines, the jump length density on the fixed domain is obtained from probability conservation:
\begin{equation}
\lambda(x,x',t)\, dx'=\lambda(y,y')\,dy',
\end{equation}
implying,
\begin{equation}
\label{rel-lamb-lamb-bar}
\lambda(x,x',t)=\bar \nu(x',t) \bar \lambda(x,x',t).
\end{equation}
On the fixed domain we now have the arrival rate probability density, for arriving at $x$ at time $t$,
\begin{equation}\label{eq_fluxqx}
\bar q(x,t|x_0,0)=\bar q_0(x,t|x_0,0) +\int_0^{L_0} \lambda(x,x',t)\int_0^t \psi(t-t')\bar\sigma(x',t,t')\bar q(x',t'|x_0,0)  dt' dx'.
\end{equation}
and
\begin{equation}\label{eq_rhotemp}
\bar{\rho}(x,t|x_0,0)=\int_0^t \Phi(t-t')\bar{\sigma}(x,t,t')\bar{q}(x,t'|x_0,0)dt',
\end{equation}
where we have defined
\begin{equation}
\bar{\sigma}(x,t,t')=e^{-\int_{t'}^t \bar{\mu}(x,s)ds}\bar{\theta}(x,t,t')=\frac{\bar \nu(x,t')}{\bar \nu(x,t)}\,\bar{\theta}(x,t,t')
\end{equation}
with $\bar\theta(x,t,t')=\theta(y,t,t')$.

To derive the evolution equation for $\bar \rho$, we differentiate Eq.~\eqref{eq_rhotemp} with respect to time.
This is complicated by the discontinuity in the arrival rate density at $t=0$ \cite{ADH2013mmnp}.
Following \cite{ADH2013mmnp}, we write
\begin{equation}\label{qqplus}
\bar q(x,t|x_0,0)=\delta(x-x_0)\delta(t-0^+)+\bar q^+(x,t|x_0,0)
\end{equation}
and then
\begin{equation}\label{eq_evo}
\bar\rho(x,t|x_0,0)=\Phi(t)\bar\sigma(x,t,0)\delta_{x,x_0}+\int_0^t\Phi(t-t')\bar{\sigma}(x,t,t')\bar{q}^+(x,t'|x_0,0)dt',
\end{equation}
where $\bar q^+(x,t|x_0,0)$ is right continuous at $t=0$.
We now use Leibniz rule, and the results $\Phi(0)=1$,  $\bar\sigma(x,t,t)=1$ and $ \partial\Phi(t-t')/\partial t=-\psi(t-t')$ to differentiate under the integral sign, arriving at
\begin{eqnarray}
\frac{\partial \bar{\rho}(x,t|x_0,0)}{\partial t}&=&-\psi(t)\bar\sigma(x,t,0)\delta_{x,x_0}
-\Phi(t)\left(\bar\mu(x,t)+\bar\omega(x,t)\right)\bar\sigma(x,t,0)\delta_{x,x_0}\nonumber\\
& &+\bar q^+(x,t)|x_0,0)
-\int_0^t\psi(t-t')\bar\sigma(x,t,t')\bar q^+(x,t'|x_0,0)\, dt'\nonumber\\
& &-\int_0^t\Phi(t-t')(\bar\mu(x,t)+\bar\omega(x,t))\bar\sigma(x,t,t')\bar q^+(x,t'|x_0,0)\, dt'.
\end{eqnarray}
This can be simplified using Eqs.~\eqref{eq_fluxqx}, \eqref{eq_evo}, \eqref{eq_rhotemp}, \eqref{qqplus} to arrive at
\begin{eqnarray} \label{eq_evo1}
\frac{\partial \bar{\rho}(x,t|x_0,0)}{\partial t}&=&\int_0^{L_0} \lambda(x,x',t)\int_0^t \psi(t-t')\bar\sigma(x',t,t')\bar q(x',t'|x_0,0)  dt' dx'\nonumber\\
& &-\int_0^t\psi(t-t')\bar\sigma(x,t,t')\bar q(x,t'|x_0,0)\, dt'\nonumber\\
& &-\left(\bar{\mu}(x,t)+\bar{\omega}(x,t)\right)\bar{\rho}(x,t|x_0,0).
\end{eqnarray}
We now wish to replace the terms involving $q$ with terms involving $\rho$.
Without loss of generality we can define a kernel $K(t-t')$ such that
\begin{equation}\label{kernel}
\int_0^t \psi(t-t')\bar q(x,t'|x_0,0)\bar\sigma(x,t,t')\, dt'
=\int_0^t K(t-t')\bar \rho(x,t'|x_0,0)\bar\sigma(x,t,t')\, dt',
\end{equation}
and then we can write Eq.~\eqref{eq_evo1} as
\begin{eqnarray} \label{eq_evo2}
\frac{\partial \bar{\rho}(x,t|x_0,0)}{\partial t}&=&\int_0^{L_0} \lambda(x,x',t)\int_0^t K(t-t')\bar\sigma(x',t,t')\bar \rho(x',t'|x_0,0)  dt' dx'\nonumber\\
& &-
\int_0^t K(t-t')\bar\sigma(x,t,t')\bar \rho(x,t'|x_0,0)\, dt'\nonumber\\
& &-\left(\bar{\mu}(x,t)+\bar{\omega}(x,t)\right)\bar{\rho}(x,t|x_0,0).
\end{eqnarray}
Using the semi-group property
\begin{equation}
\bar\sigma(x,t,0)=\bar\sigma(x,t,t')\bar\sigma(x,t',0)
\end{equation}
in Eq.~\eqref{kernel}, and taking the Laplace transform $\mathcal{L}[.]$ with respect to time in the resulting equation, we obtain
\begin{equation}\label{result1}
\mathcal{L}[\psi(t)]\mathcal{L}[\frac{\bar q(x,t|x_0,0)}{\bar \sigma(x,t,0)}]=\mathcal{L}[K(t)]\mathcal{L}[\frac{\bar \rho(x,t|x_0,0)}{\bar \sigma(x,t,0)}].
\end{equation}
On the other hand, we can also divide both sides of Eq.~\eqref{eq_rho0} by $\bar\sigma(x,t,0)$ and take the Laplace transform to find
\begin{equation}\label{result2}
\mathcal{L}[\frac{\bar \rho(x,t|x_0,0)}{\bar \sigma(x,t,0)}]=\mathcal{L}[\Phi(t)]\mathcal{L}[\frac{\bar q(x,t|x_0,0)}{\bar \sigma(x,t,0)}].
\end{equation}
Comparing Eq.~\eqref{result1} and Eq.~\eqref{result2}, we obtain the result
\begin{equation}
\label{memoryk}
K(t)=\mathcal{L}^{-1}\left[\frac{\mathcal{L}\left[\psi(t)\right]}{\mathcal{L}\left[\Phi(t)\right]}\right].
\end{equation}

\subsection{Ensemble of Particles with Birth and Death Probabilities} \label{sec_der2}

We now consider an ensemble of particles composed of individual particles that are created at particular locations, undergo random walks, and are annihilated at other locations. We hereafter assume that newborn particles are created with zero age, i.e., their ``internal clock'' used as a reference for the waiting time distribution is set to zero. Let us denote by  $\chi(y,t) dt$ the probability of a particle being created at $y$ during the interval $[t,t+dt]$.
The ensemble density of particles at location $y$ at time $t$, found by summing over all possible starting points $y_0$ and birth times $t_0$, is then given by
\begin{equation} \label{eq_umap}
u(y,t)=\int_0^{L(t)} \int_0^{t}\rho(y,t|y_0,t_0)\chi(y_0,t_0)dt_0\, dy_0.
\end{equation}
The case of a single particle initially located at $y_i$ which gives rise to no offspring is recovered by setting $\chi(y_0,t_0)=\delta(y_i-y_0)\delta(t_0-0^+)$.

Again we find it convenient to consider the auxiliary system on the fixed domain using the mapping given by Eq.~\eqref{eq_rhomap}, and the auxiliary function definitions
\begin{equation} \label{eq_uetamap}
\bar u(x,t)=u(y,t), \quad\,\mbox{and}\quad \hspace{15pt} \bar\chi(x,t)=\chi(y,t).
\end{equation}
The ensemble density of particles for the auxiliary system on the fixed initial domain is then
\begin{equation}\label{ensemble1}
\bar u(x,t)=\int_0^{t}\int_0^{L_0} \bar \nu (x_0,t) \bar\rho(x,t|x_0,t_0)\bar\chi(x_0,t_0)dx_0 dt_0.
\end{equation}
The master equation for the ensemble is found by differentiating with respect to time. In this way we find
\begin{equation} \label{diff-int-eq-u}
\frac{\partial \bar u(x,t)}{\partial t}=\int_0^t \int_0^{L_0} \bar \nu (x_0,t) \frac{\partial \bar\rho(x,t|x_0,t_0)}{\partial t}\bar\chi(x_0,t_0)dx_0  dt_0+\int_0^{L_0}\bar \nu (x_0,t)\bar\rho(x,t|x_0,t)\bar\chi(x_0,t)dx_0.
\end{equation}
This can be simplified by noting that
\begin{equation}
 \chi(y,t)=\int_0^{L(t)}\rho(y,t|y_0,t)\chi(y_0,t)dy_0,
\end{equation}
and thus
\begin{equation}
\bar\chi(x,t)=\int_0^{L_0}\bar \nu (x_0,t) \bar\rho(x,t|x_0,t)\bar\chi(x_0,t)dx_0.
\end{equation}
We further replace the derivative in Eq.~\eqref{diff-int-eq-u} using the single particle master equation Eq.~\eqref{eq_evo2}, together with Eq.~\eqref{ensemble1}, so that
\begin{equation}
\label{ensemble3}
\begin{split}
\frac{\partial \bar u(x,t)}{\partial t}&=\int_0^{L_0} \lambda(x,x',t)\int_0^t K(t-t') \bar u(x',t') \bar\sigma(x',t,t') dt'dx'\\&-\int_0^t K(t-t')\bar u(x,t')\bar\sigma(x,t,t')\, dt'-\left(\bar\mu(x,t)+\bar\omega(x,t)\right)\bar u(x,t)+\bar\chi(x,t).
\end{split}
\end{equation}
It is worth noting that in the limit of a non-growing domain $\bar \mu(x,t)\equiv 0$, one recovers results previously known from the literature.  E.g., taking $\bar \theta(x,t,t')\equiv \exp{[-\int_{t'}^t r_-(\bar{u}(x,s)) \, ds]}$ and $\bar\chi(x,t)\equiv r_+(\bar{u}(x,t))\,\bar{u}(x,t)$ leads to Eq.~(26) in Ref.~\cite{F2010} upon performing the appropriate changes in notation.

\section{Diffusion limits of auxiliary CTRW master equations}
\label{secDifLimAuxCTRW}

In this section we consider the governing equations in the diffusion limit of the master equations for CTRWs on an evolving domain. The diffusion limit requires a simultaneous limit where length scales and time scales approach zero without introducing singularities. The details of this depend on the details of the jump length and waiting time densities.
For  CTRWs on fixed domains a jump length density for unbiased jumps of fixed length, can result in standard diffusion if the waiting time density is an exponential, or subdiffusion if the waiting time density decays as a power law in time.
We now consider these possibilities on the growing domain.

The jump length density for unbiased jumps of fixed length on the growing domain can be written as
\begin{equation}
\label{jumpy}
\lambda(y,y')=\frac{1}{2}\left(\delta(y-\Delta y-y')+\delta(y+\Delta y-y')\right),
\end{equation}
where $\Delta y$ is a fixed length interval on the growing domain.
The master equations that we derived above describe the evolution of an auxiliary CTRW on the fixed domain; thus we need to map the jump density on the growing domain, Eq.~\eqref{jumpy}, to a jump density on the fixed domain.
The end positions  $y-\Delta y$ and $y+\Delta y$ after a jump are related to their corresponding positions $x-\epsilon^-$ and $x+\epsilon^+$ in the fixed domain by Eq.~\eqref{ymux}, i.e., $y-\Delta y=\bar g(x-\epsilon^-,t)$ and  $y+\Delta y=\bar g(x+\epsilon^+,t)$. Therefore, the jump density in the fixed domain is just
\begin{equation}
\label{eq_jump-dens-x}
\lambda(x,x',t)=
\frac{1}{2}\left(\delta(x-\epsilon^- -x')
+\delta(x+\epsilon^+ -x')\right).
\end{equation}
After replacing the jump density in Eq.~\eqref{ensemble3} with Eq.~\eqref{eq_jump-dens-x} we obtain
\begin{equation}
\begin{split}
\frac{\partial \bar u(x,t)}{\partial t}&=\frac{1}{2}\int_0^t K(t-t') \bar u(x-\epsilon^-,t') \bar\sigma(x-\epsilon^-,t,t') dt'\\
&+\frac{1}{2}\int_0^t K(t-t') \bar u(x+\epsilon^+,t') \bar\sigma(x+\epsilon^+,t,t') dt'\\
&-\int_0^t K(t-t')\bar u(x,t')\bar\sigma(x,t,t')\, dt'\\&-\left(\bar\mu(x,t)+\bar\omega(x,t)\right)\bar u(x,t)+\bar\chi(x,t).\label{masterx}
\end{split}
\end{equation}

In order to advance further, one now needs explicit expressions for $\epsilon^\pm$ in terms of $\Delta y$. While we have not been able to write down an explicit expression for the step sizes $\epsilon^+$ and $\epsilon^-$, it is straightforward to obtain the approximations (see the Appendix)
\begin{equation}
\label{epspm}
\epsilon^{\pm}=\frac{\Delta y}{\bar \nu(x,t)}\mp\frac{1}{2\bar \nu(x,t)^2}\left(\int_0^t \frac{\partial \bar\mu(x,s)}{\partial x}\, ds\right) \Delta y^2+O(\Delta y^3).
\end{equation}
Using the relations~\eqref{epspm}, we can perform a Taylor expansion of the functions in Eq.~\eqref{masterx} around the point $x$, retaining terms up to order $\Delta y^2$. This results in
 \begin{equation} \label{eq_amq_tay}
\begin{split}
\frac{\partial \bar u(x,t)}{\partial t}=& \frac{\Delta y^2}{2\bar \nu(x,t)^2}\left(\frac{\partial^2}{\partial x^2}\int_0^t K(t-t')\bar u(x,t')\bar\sigma(x,t,t')dt'\right.\\&\left.-\left(\int_{0}^{t}\frac{\partial \bar\mu(x,s)}{\partial x}ds\right)\frac{\partial}{\partial x}\int_0^t K(t-t')\bar u(x,t')\bar\sigma(x,t,t')dt'\right)\\&-\left(\bar\mu(x,t)+\bar\omega(x,t)\right) \bar u(x,t)+\bar\chi(x,t)+O(\Delta y^3).
\end{split}
\end{equation}

\subsection{Standard Diffusion}
In the theory of CTRWs it is well known that the standard diffusion equation can be derived from the diffusion limit of the master equation for a CTRW with nearest neighbor steps and an exponential waiting time density \cite{MK2000},
\begin{equation}
\psi(t)=\frac{1}{\tau}e^{-\frac{t}{\tau}}.
\end{equation}
The memory kernel given by Eq.~\eqref{memoryk} can readily be evaluated in this case yielding
 \begin{equation}
 K(t)=\frac{1}{\tau}\delta(t).
 \end{equation}
 Substituting this memory kernel into the  master equation for the auxiliary CTRW, Eq.~\eqref{masterx}, we obtain the result
  \begin{equation}
  \label{standard}
\begin{split}
\frac{\partial \bar u(x,t)}{\partial t}=& \frac{\Delta y^2}{2\tau \bar \nu(x,t)^2}\left(\frac{\partial^2}{\partial x^2}\bar u(x,t)\bar\sigma(x,t,t)\right.\\&\left.-\left(\int_{0}^{t}\frac{\partial \bar\mu(x,s)}{\partial x}ds\right)\frac{\partial}{\partial x}\bar u(x,t)\bar\sigma(x,t,t)\right)\\&-\left(\bar\mu(x,t)+\omega(x,t)\right) \bar u(x,t)+\bar\chi(x,t)+O(\Delta y^3).
\end{split}
\end{equation}
Finally we consider the diffusion limit, $\Delta y\to 0$ and $\tau\to 0$ with
\begin{equation}
D=\lim_{\Delta y, \tau \to 0}\frac{\Delta y^2}{2\tau},
\end{equation}
and note that $\sigma(x,t,t)=1$, to obtain
 \begin{equation} \label{standard-diff}
\begin{split}
\frac{\partial \bar u(x,t)}{\partial t}=& \frac{D}{\bar \nu(x,t)^2} \left(\frac{\partial^2}{\partial x^2}\bar u(x,t)
-\left(\int_{0}^{t}\frac{\partial \bar\mu(x,s)}{\partial x}ds\right)\frac{\partial}{\partial x}\bar u(x,t)\right)\\&-\left(\bar\mu(x,t)+\bar\omega(x,t)\right) \bar u(x,t)+\bar\chi(x,t).
\end{split}
\end{equation}
As it turns out, the equation for the ensemble density of particles $u(x,t)$ on the fixed domain is significantly simpler. Indeed, from Eq.~(\ref{eq_rhomap}) one finds
\begin{equation}
u(x,t) = \bar \nu(x,t)  \bar u(x,t),
\end{equation}
and then
\begin{equation}
\label{u-standard-diff}
\frac{\partial u(x,t)}{\partial t}=D \frac{\partial}{\partial x} \, \frac{1}{\bar \nu(x,t)}\, \frac{\partial}{\partial x} \, \frac{1}{\bar \nu(x,t)}\,u(x,t)-\bar\omega(x,t) u(x,t)+\bar\chi(x,t) \bar \nu(x,t).
\end{equation}
The first term on the right-hand-side (rhs) can be interpreted as a net probability flux, whose divergence accounts for the time change of the particle concentration inside the interval $[x,x+dx]$ in the absence of chemical reactions. In this latter case, one has the following continuity equation:
\begin{equation}
\label{cont-eq}
\frac{\partial u(x,t)}{\partial t}= D \frac{\partial}{\partial x} \,  \frac{1}{\bar \nu(x,t)}\frac{\partial}{\partial x}\, \frac{1}{\bar \nu(x,t)} u(x,t).
\end{equation}
The same equation holds if $u(x,t)$ is replaced with a density $\rho(x,t|x_0,0)$ referring to a deterministic initial condition. Note that in the case of a uniformly evolving domain $\bar \nu \equiv \nu(t)$, one obtains the diffusion equation
\begin{equation} \label{cont-eq-uniform}
\frac{\partial u(x,t)}{\partial t}= \frac{D}{\nu(t)^2} \frac{\partial^2 u(x,t)}{\partial x^2} ,
\end{equation}
a result already known from previous works (it follows e.g. by taking $v=0$ in Eq. (36) of Ref. \cite{YAE2016}). For $\bar \mu>0$, say, the shortening of the jump lengths on the fixed domain due to the growth process is described by a time-dependent effective diffusion coefficient which decreases as the inverse of the squared scale factor. The time dependence of the diffusion coefficient can be eliminated by introducing a new Brownian conformal time $\tau(t) \equiv \int_0^t \nu(s)^{-2} \, ds$ \cite{YAE2016}.

\subsubsection{Moments of $u(x,t)$ for short times}
\label{secMomND}

The analytical solution of Eq.~\eqref{u-standard-diff}  is not easy to obtain in general. An exception is the case of homogeneous expansion where $\bar \nu \equiv \nu(t)$. Fortunately, in more general cases, some useful information can still be  extracted directly from Eq.~\eqref{u-standard-diff}. In what follows, we will show how to obtain the short time behavior of the moments of $u(x,t)$ in a systematic way.  We will illustrate the procedure for the case where $\bar \mu(x,t)=\mu_0 x^2$, i.e., for $\bar \nu(x,t)=\exp[\mu_0 x^2 t]$, but the procedure can readily be carried out for other forms of the local growth function. For further simplicity, we will assume that there are no reactions, i.e., $\omega(x,t)=\bar \chi=0$.

We start by inserting the short-time power expansion of $\bar \nu(x,t)$ into Eq.~\eqref{u-standard-diff}, which yields
\begin{equation}
\label{infinite-hierarchy}
\frac{\partial u}{\partial t}=D\frac{\partial u^2}{\partial x^2}+ D\sum_{m=1}^\infty \sum_{r=0}^2 (-1)^m c_{m,r}  \mu_0^m t^m x^{2m+r-2}   \frac{\partial^r u}{\partial x^r}
\end{equation}
with $c_{1,0}=2$, $c_{1,1}=6$, $c_{1,2}=2$, $c_{2,0}=12$, etc.   Multiplying this equation by $x^n$, integrating the resulting equation over the whole interval, and assuming that $x^m u$ and $x^m \partial_x u$ are negligible for sufficiently large values of $x$,  one finds
\begin{equation}
\label{momn}
\frac{d \langle x^n \rangle}{dt}=D \sum_{m=0}^\infty a_m n(n+m-1) (\mu_0 t)^m \langle x^{n+2m} \rangle
\end{equation}
with $a_0=1$, $a_1=2$, $a_2=2$, $a_3=-4/3$, etc.  In particular, one has
\begin{align}
\label{mom1}
\frac{ d \langle x \rangle}{d t}&=D \left[
-2\mu_0t \langle x \rangle
+4 \mu_0^2 t^2 \langle x^{3} \rangle
-4 \mu_0^3 t^3 \langle x^{5} \rangle+\cdots
 \right], \\
\label{mom2}
\frac{d \langle x^2 \rangle}{d t}&=D \left[
2
-8 \mu_0 t \langle x^{2} \rangle
+12 \mu_0^2 t^2 \langle x^{4} \rangle+\cdots
 \right], \\
\label{mom3}
\frac{d \langle x^3 \rangle}{d t}&=D \left[
 6 \langle x \rangle
-18 \mu_0 t \langle x^{3} \rangle+ \cdots
 \right],\\
\label{mom4}
\frac{d \langle x^4 \rangle}{d t}&=D \left[
12  \langle x^{2} \rangle
-32 \mu_0 t \langle x^{4} \rangle+\cdots
 \right].
\end{align}
This non-closed hierarchy of equations can be solved iteratively to increasing order of powers of $t$.  For example, assume that the initial distribution of particles $u(x,0)$ has non-zero moments: $\langle x^n(0)\rangle\equiv \langle x^n_0\rangle\neq 0$.
Then, one sees that the rhs of  Eq.~\eqref{mom1} is  $ d \langle x \rangle/d t=
-2D\mu_0t \langle x_0 \rangle+O(t^2)$ because $\langle x^n\rangle=O(1)$. Thus, $\langle x \rangle= \langle x_0 \rangle-\mu_0 \langle x_0 \rangle D t^2+O(t^3)$. We can improve this approximation by noting that the next-order correction contributed by the right-hand-side of Eq.~\eqref{mom1} stems from the term proportional to $\langle x^3\rangle$ and is quadratic in time. This is because, from Eq.~\eqref{mom3}, we know that $\langle x^3 \rangle\approx \langle x^3_0 \rangle+O(t)$. Therefore,  to order $t^2$, Eq.~\eqref{mom1} can be written as follows:
\begin{equation}
\frac{d \langle x^3 \rangle}{d t}=D \left[
-2\mu_0t \langle x_0 \rangle
+4 \mu_0^2 t^2 \langle x^{3}_0 \rangle  \right] +O(t^3).
\end{equation}
Thus,
\begin{equation}
\label{mom1sO3}
\langle x \rangle=
\langle x_0 \rangle
-\mu_0 \langle x_0 \rangle D t^2
+\frac{4}{3} \mu_0^2 \langle x^3_0 \rangle D t^3
+O(t^4).
\end{equation}
Other moments can be evaluated with the same procedure.  For example,  for the third-order moment, one finds
\begin{equation}
\label{mom3sO2}
\langle x^3 \rangle=
\langle x^3_0 \rangle
+6 \langle x_0 \rangle D t
-9 \mu_0 \langle x^3_0 \rangle  D t^2
 +O(t^3).
\end{equation}
The result~\eqref{mom1sO3} for $\langle x \rangle$ can be further improved by inserting Eqs.~\eqref{mom1sO3} and Eq.~\eqref{mom3sO2} into Eq.~\eqref{mom1}, and then taking into account that $\langle x^5 \rangle\approx \langle x^5_0 \rangle+O(t)$.  This iterative procedure works also for even-order moments. In this way one finds
\begin{equation}
\label{mom2sO2}
\langle x^2 \rangle=
\langle x^2_0 \rangle
+2  D t
-4 \mu_0 \langle x^2_0 \rangle  D t^2
 +O(t^3)
\end{equation}
and
\begin{equation}
\label{mom4sO1}
\langle x^4 \rangle=
\langle x^4_0 \rangle
+12 \langle x^2_0\rangle D t
 +O(t^2).
\end{equation}
Improved expressions with an additional corrective term are given by
\begin{align}
\langle x \rangle=&
\langle x_0 \rangle
-\mu_0 \langle x_0 \rangle D t^2
+\frac{4}{3} \mu_0^2 \langle x^3_0 \rangle D t^3
+ \left(\frac{13}{2} D  \langle x_0 \rangle - \mu_0 \langle x^5_0 \rangle\right) D \mu_0^2 t^4
+O(t^5),   \label{mom1sO4} \\
\langle x^2 \rangle=&
\langle x^2_0 \rangle
+2  D t
-4 \mu_0 \langle x^2_0 \rangle  D t^2
+ \left(4  \mu_0 \langle x^4_0 \rangle -\frac{16}{3} D \right) \mu_0 D t^3
 +O(t^4), \label{mom2sO3}\\
 \langle x^3 \rangle=&
\langle x^3_0 \rangle
+6 \langle x_0 \rangle D t
-9 \mu_0 \langle x^3_0 \rangle  D t^2
+ \left(8 \mu_0  \langle x^5_0 \rangle-38 D \langle x_0 \rangle \right)\mu_0 D t^3
 +O(t^4), \label{mom3sO4}\\
 \langle x^4 \rangle=&
\langle x^4_0 \rangle
+12 \langle x^2_0\rangle D t
+\left(12 D - 16  \mu_0 \langle x^4_0 \rangle\right) D t^2
 +O(t^3).  \label{mom4sO2}
\end{align}

By inspection, one easily notes that, at least up to the fourth-order moment, the leading correction to $\langle x^m\rangle$ arising from the domain growth takes the form $-m^2 \mu_0 \langle x_0^m \rangle Dt^2$.  We note the important role of the initial condition $x_0$, as opposed to the case of a uniformly evolving domain. In the case of the first-order moment, when $\mu_0>0$, the term $-\mu_0 \langle x_0\rangle Dt^2$ reflects an accelerated motion towards the origin as a result of the confining effect of the domain growth in $x$-space (for $\mu_0<0$, this motion is directed away from the origin).

\begin{figure}
\includegraphics[width=0.49\textwidth]{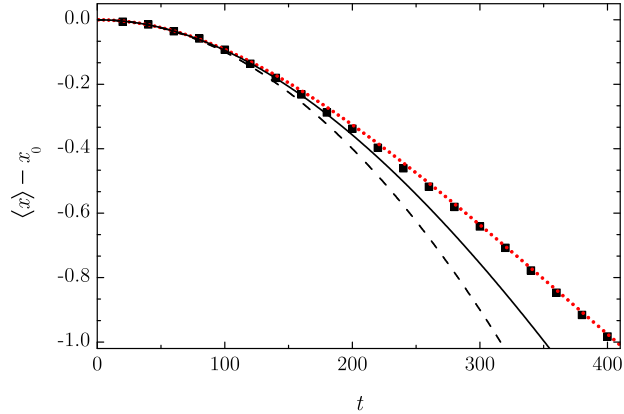}
\caption{$\langle x\rangle - x_0$ vs. time for  $\mu(x,t)=\mu_0 x^2$,  $u(x,0)=\delta(x-x_0)$, $x_0=20$, $D=1/2$ and $\mu_0=10^{-6}$. The squares are simulation results. The dotted line corresponds to results obtained from the numerical solution of Eq.~\eqref{cont-eq}. The broken and solid lines are the analytical expressions to order $t^2$ and order $t^3$, respectively, given in Eq.~\eqref{mom1sO3}.}
\label{fig:mom1x}
\end{figure}

\begin{figure}
\includegraphics[width=0.49\textwidth]{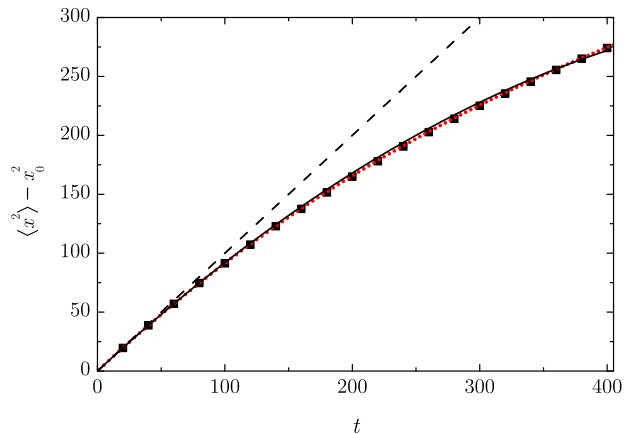}
\caption{$\langle x^2\rangle - x_0^2$ vs. time for  $\mu(x,t)=\mu_0 x^2$, $u(x,0)=\delta(x-x_0)$ with $x_0=20$, $D=1/2$ and $\mu_0=10^{-6}$. The squares are simulation results. The dotted line corresponds to results obtained from the numerical solution of Eq.~\eqref{cont-eq}. The broken and solid lines are the analytical expressions to order $t$  and order $t^2$, respectively, given in Eq.~\eqref{mom2sO2}. Note that the expression of order $t$ is just the well-known expression for a static domain (case $\mu_0=0$).}
\label{fig:mom2x}
\end{figure}

 In Fig.~\ref{fig:mom1x} and Fig.~\ref{fig:mom2x} we compare the  results from Eq.~\eqref{mom1sO3} and Eq.~\eqref{mom2sO2}, respectively, with estimates of the moments  obtained from simulations and from numerical solutions of Eq.~\eqref{cont-eq}. The simulation method for non-uniform domain evolution is a straightforward generalization of that described in Sec.~III.E  of Ref.~\cite{VY2018} for the case of a uniformly evolving domain.  We have taken the waiting time pdf and the jump length pdf used therein to perform the simulations.

For the special case where all the particles are initially placed at $x=0$, one has $\langle x^n_0\rangle=0$ for $n\ge 1$, $\langle x^n\rangle=0$ for $n=\text{odd}$, and Eqs.~\eqref{mom2sO3} and~\eqref{mom4sO2} become
\begin{align}
\langle x^2 \rangle=&
2  D t
-\frac{16}{3}  \mu_0 D^2 t^3
 +O(t^5), \label{mom2sO3delta0}\\
 \langle x^4 \rangle=&
12 D^2  t^2
 +O(t^4).  \label{mom4sO2delta0}
\end{align}

The above findings highlight yet again the importance of the initial condition, to the extent that the time dependence of the leading correction to the second- and fourth-order moments is different than that obtained for the case $x_0\neq 0$. For example, the short-time correction to leading order is cubic in the present case, and therefore weaker than the quadratic dependence obtained when $x_0=0$ [cf. Eq.~\eqref{mom2sO3}].

\begin{figure}
\includegraphics[width=0.49\textwidth]{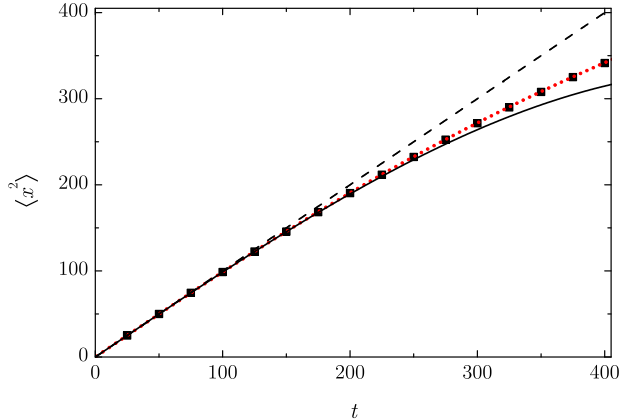}
\caption{$\langle x^2\rangle$ vs. time for  $\mu(x,t)=\mu_0 x^2$, $u(x,0)=\delta(x)$,  $D=1/2$ and $\mu_0=10^{-6}$. The squares are simulation results. The dotted line corresponds to results obtained from the numerical solution of Eq.~\eqref{cont-eq}. The broken and solid lines are the analytical expressions to order $t$ (or equivalently, for a static domain)  and order $t^3$, respectively, given in Eq.~\eqref{mom2sO3delta0}. }
\label{figmom2xFelipe}
\end{figure}

Comparisons between moment calculations and simulations for  $\langle x^2 \rangle$, with the initial distribution of particles given by a Dirac delta function at $x=0$,  are shown in Fig.~\ref{figmom2xFelipe}.
It is noteworthy that $\langle x^2 \rangle$ does not scale linearly in time, as anticipated on a non-evolving domain.
Note also that the moment calculations reproduce the numerical results with improved accuracy as the order is increased. For the case $x_0=0$, it is instructive to compare the result~\eqref{mom2sO3delta0} with the case of a uniform exponential growth/contraction with $\nu(t)=\mbox{exp}[\hat{\mu}_0 t]$. In this case, one has the exact result,
\begin{equation}
\langle x^2 \rangle_\text{exp}=2D\tau(t)=D\hat{\mu}_0^{-1}(1-e^{-2\hat{\mu}_0 t}),
\end{equation}
valid for all times $t$. For short times this can be expanded as
\begin{equation}
\langle x^2 \rangle_\text{exp}=2Dt-2D\hat{\mu}_0 t^2+O(t^3).
\end{equation}
The correction (positive or negative) due to the domain evolution at early times is stronger in the uniform case, since it is proportional to $t^2$.

Let us focus on the case of a growing domain. In contrast with the case  $\mu(x,t)=\mu(x)=\mu_0 x^2$, for a uniform exponential growth the particle motion on the fixed domain experiences a strong confinement already for small excursions from the origin, typically corresponding to short travel times $t$. This is due to the aforementioned effective reduction of the diffusion coefficient on the fixed domain [cf. Eq.~\eqref{cont-eq-uniform}]. Therefore, after a short time $t$, the correction to a pure diffusive motion is more important than in the case $\mu(x)\propto x^2$, where the domain growth is practically zero at short distances from the origin and $\bar{\nu}(x,t)\approx 1$ in this regime [cf. Eq.~\eqref{cont-eq}]. An analogous reasoning applies for the contracting case $\mu(x), \hat{\mu}_0<0$.

Finally, it is also worth mentioning that, in the contracting case $\mu_0<0$, it is possible to obtain a hierarchy that is valid at all times, not only in the early-time regime. To this end, one uses a modified local growth rate  $\bar{\mu}(x,t)=\mu_0 x^2/(1-\mu_0 x^2t)$, with $\mu_0<0$. This yields $\nu(x,t)^{-1}=1-\mu_0 x^2t$ for all times, whence the exact (albeit non-closed) hierarchy
\begin{equation}
\label{1sthierarchy}
\frac{d\langle x^n \rangle}{dt}=D n (n-1) \langle x^{n-2} \rangle -2n^2\mu_0 D t \,\langle x^n \rangle + n(n+1) \mu_0^2 Dt^2 \, \langle x^{n+2} \rangle,
\end{equation}
follows.

\subsubsection{Moments of $u(y,t)$ for short times}
\label{secMomNDy}

The short-time moments of $u(x,t)$ can be employed to get the short-time moments of $u(y,t)$ by expanding $y^n=\bar g(x,t)^n$. For example, to first order in $t$ one has   $y^n=x^n+n t\int_0^x \bar \mu(z,0) dz + O(t^2)$.  In particular, for $\bar\mu(x,t)=\mu_0 x^2$, one finds
\begin{subequations}
\begin{equation}
\label{short-time-y}
\langle y\rangle=\langle x\rangle +\frac{1}{3}\,\mu_0 t \langle x^3\rangle+\frac{1}{10}\,\mu_0^2 t^2 \langle x^5\rangle + O(\langle x^7\rangle t^3)
\end{equation}
 and
\begin{equation}
\label{short-time-y2}
 \langle y^2\rangle=\langle x^2\rangle +\frac{2}{3}\,\mu_0 t \langle x^4\rangle+\frac{14}{45}\,\mu_0^2 t^2 \langle x^6\rangle+ O(\langle x^8\rangle t^3).
\end{equation}
\end{subequations}
Then
\begin{equation}
\label{momy1sO2i}
\langle y \rangle=
\langle y_0 \rangle
+\frac{1}{3} \mu_0 \langle y^3_0 \rangle   t
+\left(\mu_0 \langle y_0 \rangle D +\frac{\mu_0^2}{10}  \langle y^5_0 \rangle\right)t^2
 +O(t^3)
\end{equation}
and
\begin{equation}
\label{momy2sO2i}
\langle y^2 \rangle=
\langle y^2_0 \rangle
+\left(2 D +\frac{2}{3} \mu_0 \langle y^4_0 \rangle\right)   t
+\frac{2}{45}\left(90  \mu_0  \langle y^2_0 \rangle D +7 \mu_0^2 \langle y^6_0 \rangle\right)   t^2
 +O(t^3).
\end{equation}

It is worth noting that the leading correction to the first and second-order moment has the same sign as $\mu_0$ and is linear in time; therefore, it is stronger than the quadratic correction observed in the case of the $x$-moments. In the case of the first order moment, the leading correction can be interpreted as a deterministic drift of the form $vt$, with a velocity given by $v=\mu_0 \langle y_0^2\rangle/3$. Note that the diffusivity $D$ does not appear in this term; indeed, the diffusive motion is a subleading correction to the dominant drift arising from the domain evolution.

In contrast, the leading contribution of the domain evolution to the second order moment is of the same order as the intrinsic diffusive motion, leading to an apparent diffusivity $D+\mu_0 \langle y^4_0\rangle /3$ which describes particle spreading to dominant order.

 For the case $y=0$, insertion of Eqs.~\eqref{mom2sO3delta0} and~\eqref{mom4sO2delta0} into
 Eq.~\eqref{short-time-y2} yields
 \begin{equation}
 \label{centered-y2}
\langle y^2 \rangle=2Dt+\frac{8}{3}\mu_0 D^2 t^3+O(t^5).
\end{equation}

\begin{figure}
\includegraphics[width=0.49\textwidth]{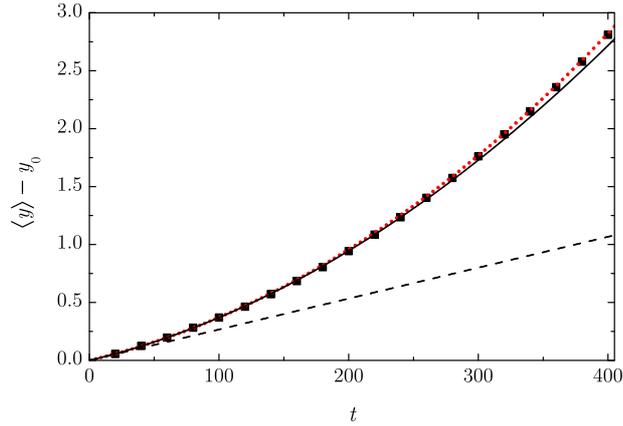}
\caption{$\langle y\rangle - y_0$ vs. time for  $\mu(x,t)=\mu_0 x^2$ and $u(y,0)=\delta(y-y_0)$ with $y_0=20$, $D=1/2$ and $\mu_0=10^{-6}$. The squares are simulation results. The dotted line is obtained from the numerical solution of Eq.~\eqref{cont-eq}. The broken and solid lines are the analytical expressions to order $t$ and order $t^2$, respectively, given in Eq.~\eqref{momy1sO2i}. }
\label{fig:mom1y}
\end{figure}

\begin{figure}
\includegraphics[width=0.49\textwidth]{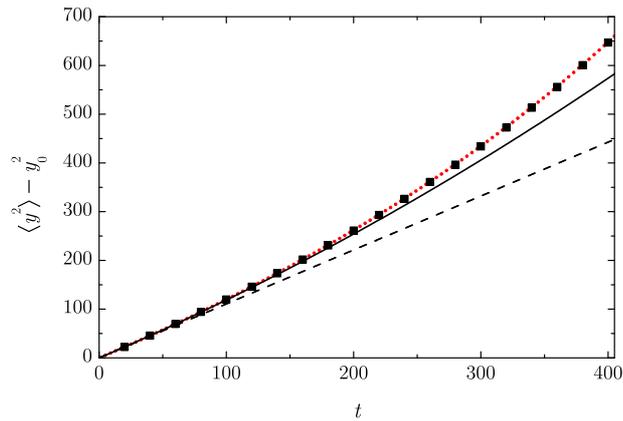}
\caption{$\langle y^2\rangle - y_0^2$ vs. time for  $\mu(x,t)=\mu_0 x^2$ and $u(y,0)=\delta(y-y_0)$ with $y_0=20$, $D=1/2$ and $\mu_0=10^{-6}$. The dotted line is obtained from the numerical solution of Eq.~\eqref{cont-eq}. The broken and solid lines are the analytical expressions to order $t$ and order $t^2$, respectively, given in Eq.~\eqref{momy2sO2i}. } \label{fig:mom2y}
\end{figure}

 In Figs.~\ref{fig:mom1y} and \ref{fig:mom2y} we compare these  results with numerical estimates of the moments  $\langle y^n\rangle$. These results have been obtained by solving  Eq.~\eqref{cont-eq} numerically. From the numerical solution $u(x,t)$ one then finds   $u(y=\bar g(x,t),t)$ and, from this function, $\langle y^n\rangle$.

\begin{figure}
\includegraphics[width=0.49\textwidth]{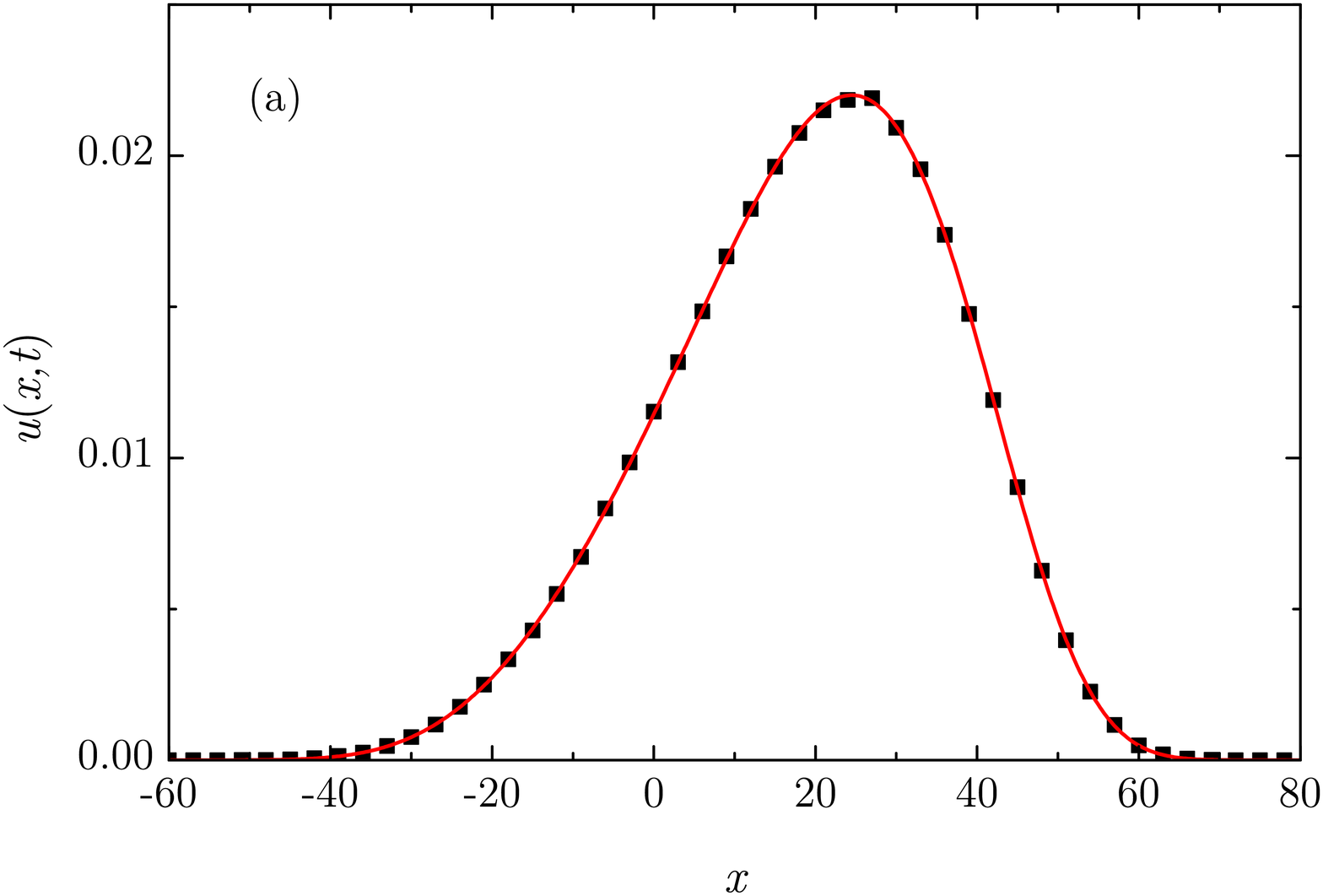}
\includegraphics[width=0.49\textwidth]{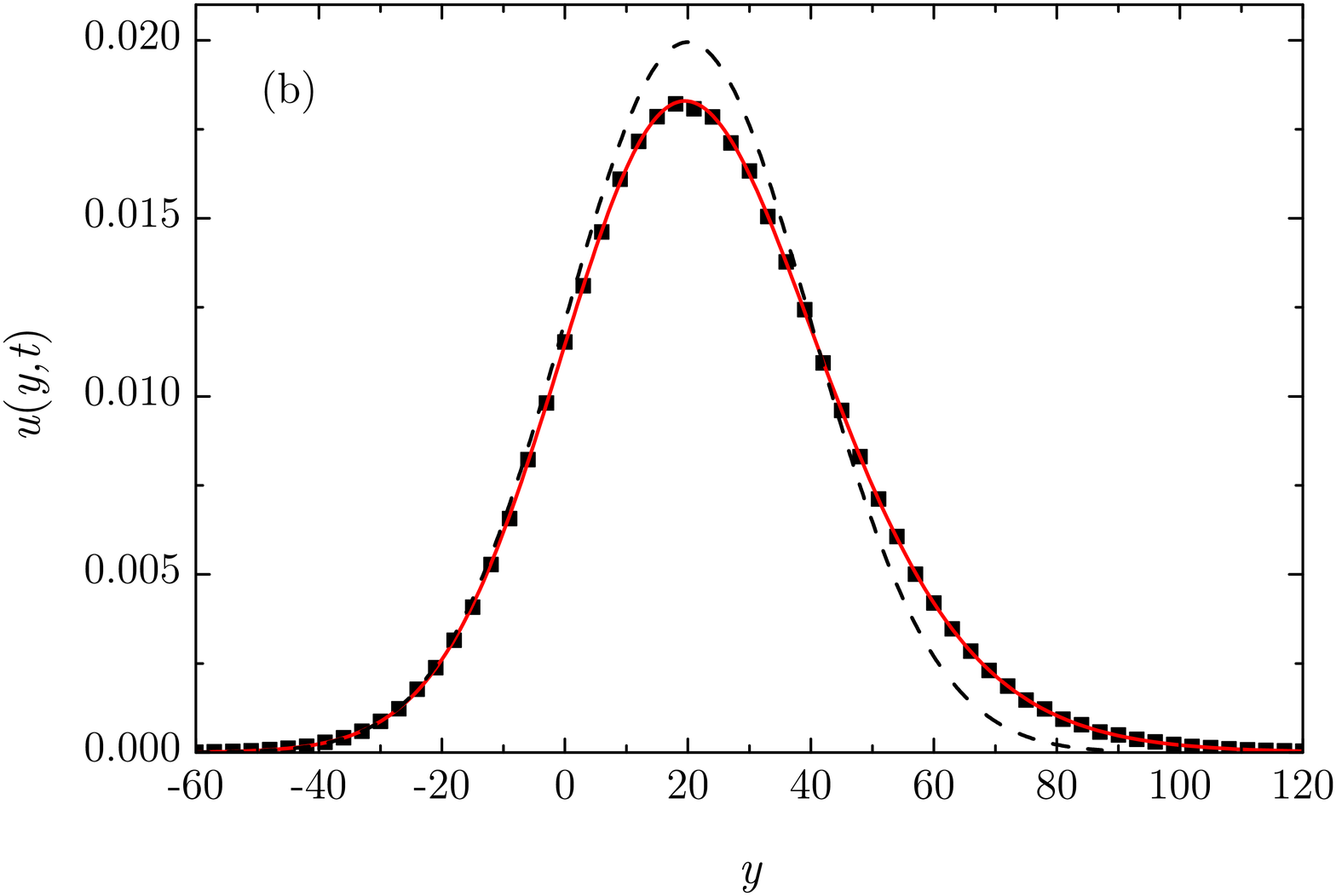}
\caption{(a) $u(x,t)$  and (b)  $u(y,t)$   for $t=400$, $D=1/2$, $\mu_0=10^{-6}$,  and $u(x,0)=\delta(x-x_0)$ with $x_0=20$. This deterministic initial condition implies  $u(x,t)=\rho(x,t)$. The solid lines represent the numerical solution of Eq.~\eqref{cont-eq}, whereas the squares are simulation results. The broken line corresponds to the solution for the static case ($\mu_0=0$).}
\label{fig:uxyt400}
\end{figure}

\begin{figure}
\includegraphics[width=0.49\textwidth]{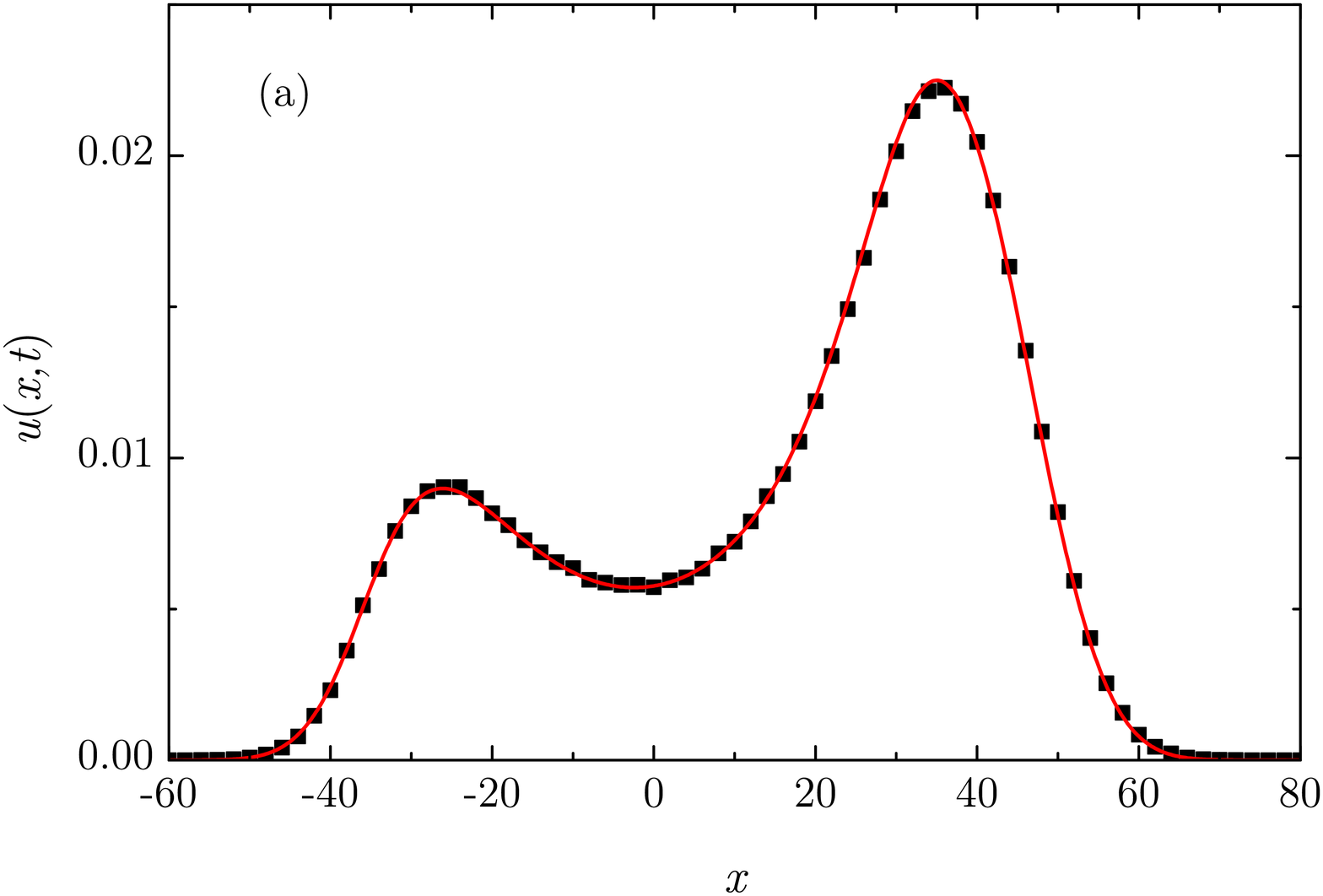}
\includegraphics[width=0.49\textwidth]{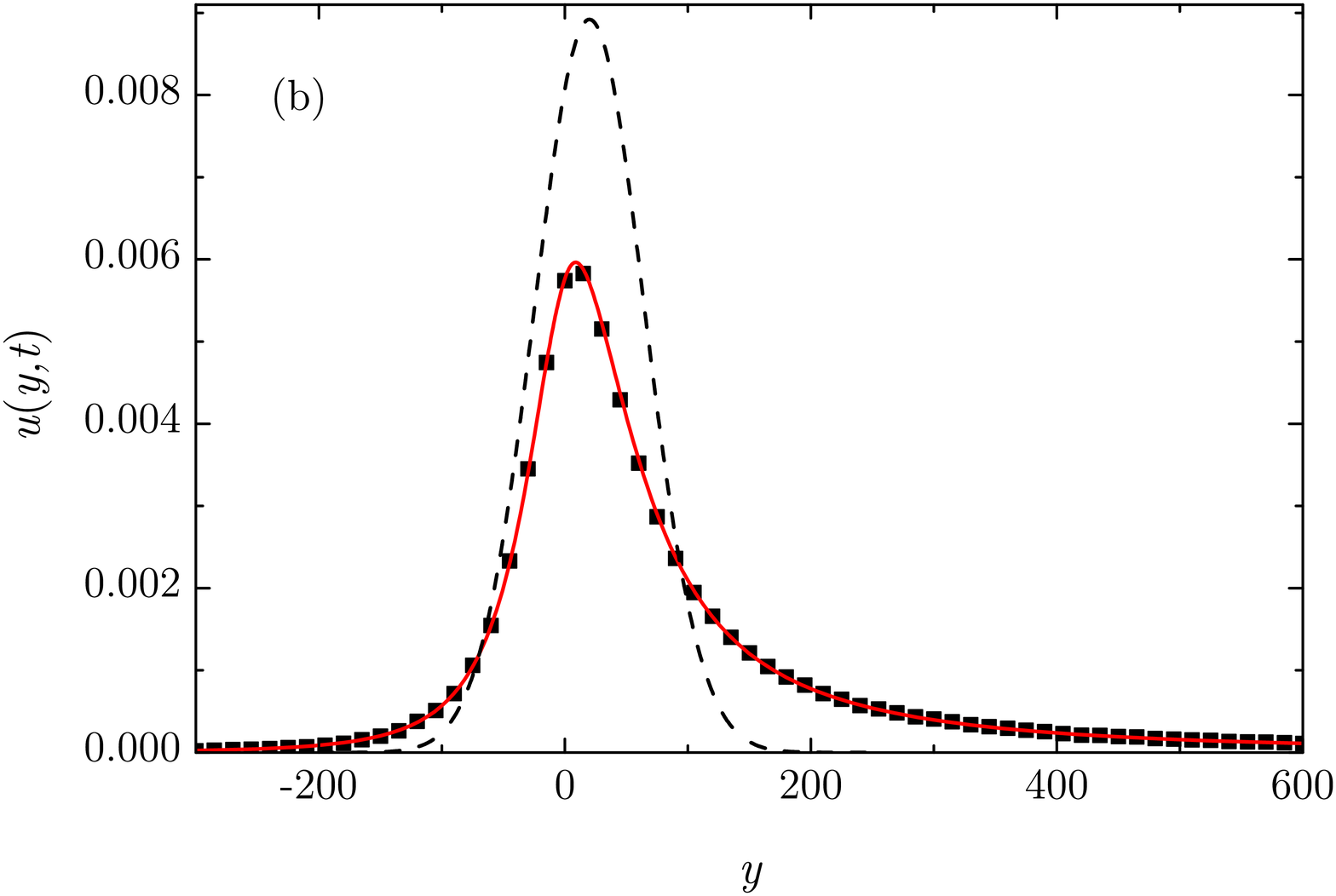}
\caption{ (a) $u(x,t)$  and (b)  $u(y,t)$   for $t=2000$, $D=1/2$, and $\mu_0=10^{-6}$ and $u(x,0)=\delta(x-x_0)$ with  $x_0=20$. The lines represent the  numerical solution of Eq.~\eqref{cont-eq}  whereas the squares are simulation results. The broken line corresponds to the solution for the static case ($\mu_0=0$).}
\label{fig:uxyt2000}
\end{figure}

As complementary information, we show the numerical plots of the $u(x,t)$ and $u(y,t)$ for $t=400$ and $t=2000$ in Figs.~\ref{fig:uxyt400} and \ref{fig:uxyt2000}. The bimodal form of $u(x,t=2000)$ shows the clear influence of the evolving domain on the distribution for the auxiliary domain.

As an aside we note that, in view of the relation
\begin{equation}
u(k,t)={\cal F}\left\{u(x,t)\right\}=\int_{-\infty}^\infty e^{-ikx} u(x,t) \,dx=\sum_{m=0}^\infty \frac{(-ik)^m}{m!} \langle x^m(t) \rangle,
\end{equation}
the obtained expressions for the moments may be used to obtain an early-time approximation for the full Fourier-transformed probability density function by truncating the above series to different orders.

Finally, we note that the results obtained for the moments $\langle x^n\rangle$ in the absence of reactions,  $  \omega=\bar\chi=0$,  can be straightforwardly extended to include a pure death process with constant rate ($\bar \omega=\omega_0$, and $\bar \chi=0$).  This is done by noting that  $\hat u (x,t)$ for this case and the corresponding solution $u(x,t)$ for the case without reactions, are related to one another by  $\hat u(x,t)= u(x,t) \exp[-\omega_0 t]$. Therefore $\langle \hat x^n \rangle= \langle   x^n \rangle \exp[-\omega_0 t]$, where $\langle \hat x^n \rangle$ and $\langle   x^n \rangle$ are the moments associated with $\hat u$ and $u$, respectively.

\subsection{Subdiffusion}
Subdiffusion can be obtained from CTRWs with a heavy-tailed power-law waiting time density \cite{MK2000}.
The Mittag-Leffler density has been widely studied in this context. This is defined as
\begin{equation}
\label{eq_psi}
\psi(t)= \frac{t^{\alpha-1}}{\tau^{\alpha}}E_{\alpha,\alpha}\left(-\left(\frac{t}{\tau}\right)^\alpha\right),
\end{equation}
where
\begin{equation}
E_{\alpha,\beta}(z)=\sum_{k=0}^\infty \frac{z^k}{\Gamma(\alpha k+\beta)}
\end{equation}
is the Mittag-Leffler function.

Rather than substitute the corresponding memory kernel directly into the master equation, we first note a number of Laplace transform properties. Using the expression $\mathcal{L}[\psi(t)]\equiv \psi(s)=1/(\tau^\alpha s^\alpha+1)$ for the Laplace transform of the density \eqref{eq_psi} in  Eq.~\eqref{memoryk}, we obtain \cite{Podlubny1999}
\begin{equation}
\mathcal{L}[K(t)]=\frac{s^{1-\alpha}}{\tau^\alpha},
\end{equation}
whence the result
\begin{equation}
\label{property1}
\mathcal{L}\int_0^t K(t-t') Y(t')\, dt'=\frac{s^{1-\alpha}}{\tau^\alpha}\mathcal{L}[Y(t)],
\end{equation}
follows by straightforward application of the convolution theorem for the Laplace transform.

We note, on the other hand, that
\begin{equation}
\label{property2}
\mathcal{L}[\, _0\mathcal{D}_t^{1-\alpha} Y(t)]=s^{1-\alpha}\mathcal{L}[Y(t)],
\end{equation}
where $\mathcal{D}_t^{1-\alpha} Y(t)$ denotes the Grünwald-Letnikov fractional derivative
of order $1-\alpha$ of the function $Y(t)$. This operator is known to be equivalent to the Riemann-Liouville fractional derivative,
\begin{equation}
\, _0 D_t^{1-\alpha} (Y(x,t))=\frac{1}{\Gamma(\alpha)}
\frac{\partial}{\partial t}\int_0^t \frac{Y(x,t')}{(t-t')^{1-\alpha}}\, dt'
\end{equation}
for sufficiently smooth functions (see e.g. Eqs. (13-15) in Ref.~\cite{LVYA2019}).

Comparing Eq.~\eqref{property1}with Eq.~\eqref{property2}, we find that for a Mittag-Leffler memory kernel one has
\begin{equation}
\int_0^t K(t-t') Y(t')\, dt'=\frac{1}{\tau^{\alpha}}\, _0\mathcal{D}_t^{1-\alpha} Y(t).
\end{equation}
Consequently, the auxiliary CTRW master equation, Eq.~\eqref{masterx}, can be written as
 \begin{equation}
\begin{split}
\frac{\partial \bar u(x,t)}{\partial t}=&
 \frac{\Delta y^2 \bar \nu(x,t)^{-2}}{2\tau^\alpha}
 \left(\frac{\partial^2}{\partial x^2}\left(\bar\sigma(x,t,0)\, _0\mathcal{D}_t^{1-\alpha}
 \frac{\bar u(x,t)}{\bar\sigma(x,t,0)}\right)\right.\\&\left.-\left(\int_{0}^{t}\frac{\partial \bar\mu(x,s)}{\partial x}ds\right)\frac{\partial}{\partial x}\left(\bar\sigma(x,t,0)\, _0\mathcal{D}_t^{1-\alpha} \frac{\bar u(x,t)}{\bar\sigma(x,t,0)}\right)\right)\\&-\left(\bar\mu(x,t)+\bar\omega(x,t)\right) \bar u(x,t)+\bar\chi(x,t)+O(\Delta y^3).
\end{split}
\end{equation}
We consider the diffusion limit, $\Delta y\to 0$ and $\tau\to 0$ with
\begin{equation}
D=\lim_{\Delta y, \tau \to 0}\frac{\Delta y^2}{2\tau^\alpha},
\end{equation}
and then
 \begin{equation}\label{sub}
\begin{split}
\frac{\partial \bar u(x,t)}{\partial t}=&
D_\alpha \bar \nu(x,t)^{-2}
 \left(\frac{\partial^2}{\partial x^2}\left(\bar\sigma(x,t,0)\, _0\mathcal{D}_t^{1-\alpha}
 \frac{\bar u(x,t)}{\bar\sigma(x,t,0)}\right)\right.\\&\left.-\left(\int_{0}^{t}\frac{\partial \bar\mu(x,s)}{\partial x}ds\right)\frac{\partial}{\partial x}\left(\bar\sigma(x,t,0)\, _0\mathcal{D}_t^{1-\alpha} \frac{\bar u(x,t)}{\bar\sigma(x,t,0)}\right)\right)\\&-\left(\bar\mu(x,t)+\bar\omega(x,t)\right) \bar u(x,t)+\bar\chi(x,t).
\end{split}
\end{equation}

Eq.~\eqref{sub} can be rewritten as
\begin{equation}\label{sub2}
\begin{split}
\frac{\partial \bar u(x,t)}{\partial t}=&
D_\alpha \bar \nu(x,t)^{-1}
\frac{\partial}{\partial x}
 \left[\bar \nu(x,t)^{-1}\frac{\partial}{\partial x}\left(\bar\sigma(x,t,0)\, _0\mathcal{D}_t^{1-\alpha} \frac{\bar u(x,t)}{\bar\sigma(x,t,0)}\right)\right]
\\
&
-\left(\bar\mu(x,t)+\bar\omega(x,t)\right) \bar u(x,t)+\bar\chi(x,t)
\end{split}
\end{equation}
or, in terms of the  ensemble density of particles on the fixed domain, $u(x,t)=\bar \nu(x,t) \bar u(x,t)$, as
\begin{equation}
\label{sub3}
\begin{split}
\frac{\partial  u(x,t)}{\partial t}=&
D_\alpha
\frac{\partial}{\partial x}\left\{
 \bar \nu(x,t)^{-1}
 \frac{\partial}{\partial x}\left[
 \bar \nu(x,t)^{-1} e^{-\int_{0}^{t}\bar\omega(x,s)ds}\, _0\mathcal{D}_t^{1-\alpha} \frac{ u(x,t)}{e^{-\int_{0}^{t}\bar\omega(x,s)ds}}\right]\right\}
\\
&
- \bar\omega(x,t) u(x,t)+\bar\chi(x,t) \bar \nu(x,t)
\end{split}
\end{equation}
where we have used the relationship
$$\bar\sigma(x,t,0)=\bar\theta(x,t,0) \bar \nu(x,t)^{-1}=e^{-\int_{0}^{t}\bar\omega(x,s)ds}
\bar \nu(x,t)^{-1}.$$
For the case $\omega(x,t)=\omega(t)$ one finds a simpler expression in terms of $u^*(x,t)=u(x,t)e^{\int_{0}^{t}\bar\omega(s)ds}$,
\begin{equation}
\label{sub4}
\begin{split}
\frac{\partial  u^*(x,t)}{\partial t}=&
D_\alpha
\left\{
\frac{\partial}{\partial x}
\bar \nu(x,t)^{-1}
 \frac{\partial}{\partial x}\left[
 \bar \nu(x,t)^{-1} \, _0\mathcal{D}_t^{1-\alpha} u^*(x,t)\right]\right\}
\\
&
+\bar\chi(x,t) \bar \nu(x,t) e^{-\int_{0}^{t}\bar\omega(x,s)ds}.
\end{split}
\end{equation}

\begin{figure}
\includegraphics[width=0.49\textwidth]{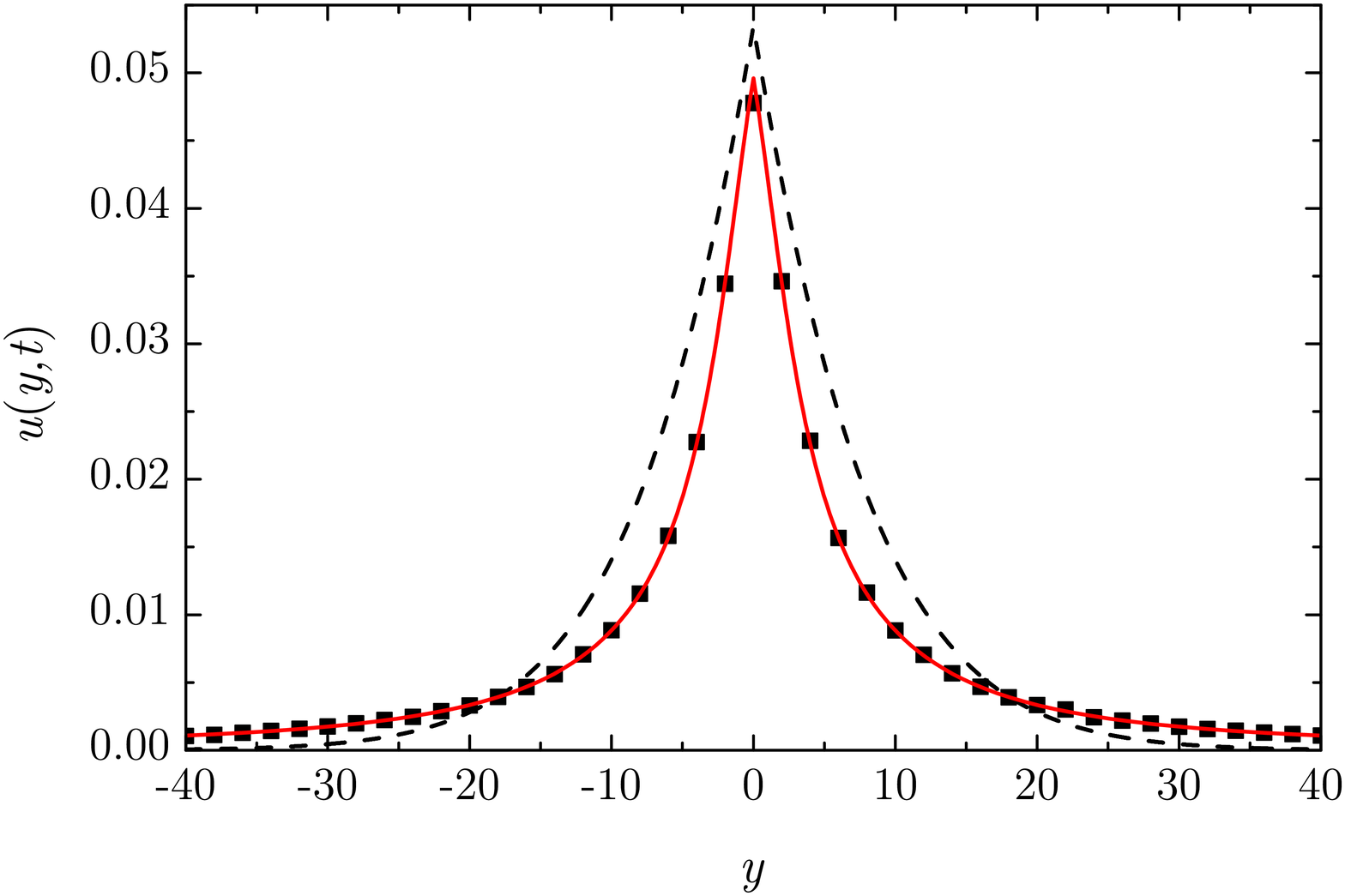}
\caption{Numerical solution $u(y,t)$ (solid line) and simulation results (squares)  for
 $\alpha=1/2$, $D_\alpha=1/2$, $\mu_0=10^{-5}$, $\omega(x,t)=\omega_0=5\times10^{-5}$ and $u(x,0)=\delta(x)$.  The broken line is the corresponding solution for the case of a static domain.
}
\label{figuxtSub}
\end{figure}

In general, Eq.~\eqref{sub3} cannot be solved exactly, but one can resort to numerical methods instead. For example, in Fig.~\ref{figuxtSub}, we show the numerical solution of Eq.~\eqref{sub3} for a particular yet representative case.
The solution has been obtained by means of a straightforward extension of the fractional Crank-Nicolson method described in Ref.~\cite{Yuste2006}.

\subsubsection{Moments of $u(x,t)$ for short times}

We proceed here in the same way as for the case of normal diffusion.  For example, assume that we have an expansion of the form $\mu(x,t)=\mu_0 x^2$, and no reactions.  Then, it is easy to see that the hierarchy of moment equations obtained for subdiffusion is recovered from that in Sec.~\ref{secMomND}, Eqs.~\eqref{momn}--\eqref{mom4},  by simply replacing $\langle x^n \rangle$ with $\, _0\mathcal{D}_t^{1-\alpha}\langle x^n \rangle$  and $D$ with $D_\alpha$.  The iterative procedure for solving this hierarchy of equations is also similar to that employed in Sec.~\ref{secMomND}.
The first two equations for odd-order moments are [cf. Eqs.~\eqref{mom1} and~\eqref{mom3}]:
\begin{align}
\label{mom1sub}
\frac{d \langle x \rangle}{d t}&=D_\alpha \left[
-2\mu_0t  \,_0\mathcal{D}_t^{1-\alpha} \langle x \rangle
+4 \mu_0^2 t^2  \,_0\mathcal{D}_t^{1-\alpha} \langle x^{3} \rangle
-4 \mu_0^3 t^3  \,_0\mathcal{D}_t^{1-\alpha} \langle x^{5} \rangle+\cdots
 \right],   \\
\label{mom3sub}
\frac{d \langle x^3 \rangle}{d t}&=D_\alpha \left[
 6  \,_0\mathcal{D}_t^{1-\alpha} \langle x \rangle
-18 \mu_0 t  \,_0\mathcal{D}_t^{1-\alpha} \langle x^{3} \rangle+ \cdots
 \right].
\end{align}
To order $t^0$ we have $\langle x^n \rangle(t) =\langle  x^n \rangle(0)\equiv \langle  x^n_0 \rangle$, and so Eq.~\eqref{mom1sub}  can be approximated, to order $t^\alpha$, by
\begin{equation}
\label{eqmom1xsubOalpha}
\frac{d\langle x \rangle}{d t}=-2\mu_0 t D_\alpha \,_0\mathcal{D}_t^{1-\alpha} \langle x_0 \rangle  +o\left( t^{\alpha}\right)=  -2\mu_0  D_\alpha \langle x_0 \rangle \frac{t^\alpha }{\Gamma(\alpha)} +o\left( t^{\alpha}\right).
\end{equation}
Here we have used the fact that   $\,_0\mathcal{D}_t^{1-\alpha}\, 1=  t^{\alpha-1}/\Gamma(\alpha)$, which is a special case of the more general result
\begin{equation}
\label{Dfrac}
\,_0\mathcal{D}_t^{1-\alpha} \,t^\beta= \frac{\Gamma(1+\beta)}{\Gamma(\alpha+\beta)}\, t^{\alpha+\beta-1}.
\end{equation}
The integration of Eq.~\eqref{eqmom1xsubOalpha} yields
\begin{equation}
\label{mom1xsubOalpha}
  \langle x \rangle = \langle x_0 \rangle -\frac{2\mu_0  \langle x_0 \rangle  }{(1+\alpha)\Gamma(\alpha)} D_\alpha  \, t^{1+\alpha} +o\left( t^{1+\alpha}\right).
  \end{equation}
Similarly, one finds
\begin{equation}
\label{mom3xsubOalpha}
  \langle x^3 \rangle = \langle x^3_0 \rangle +\frac{6 \mu_0 \langle x_0 \rangle }{ \Gamma(1+\alpha)}   D_\alpha   \, t^\alpha+o\left( t^{\alpha}\right).
\end{equation}
We can improve these approximations in an iterative way.  For example, from the approximations  in Eqs.~\eqref{mom1xsubOalpha} and ~\eqref{mom3xsubOalpha}, we find that  the equation for $d\langle x\rangle/dt$ to order $t^{1+\alpha}$, is
\begin{equation}
\label{eqmom1xsubO1palpha}
\frac{d \langle x \rangle}{d t} =
-2\mu_0  D_\alpha t \,_0\mathcal{D}_t^{1-\alpha} \langle x_0 \rangle
+4 \mu_0^2 D_\alpha t^2  \,_0\mathcal{D}_t^{1-\alpha} \langle x_0^{3} \rangle +o\left( t^{1+\alpha}\right).
\end{equation}
 Taking into account  Eq.~\eqref{Dfrac} and integrating the resulting equations, we easily find the first-order moment up to order $t^{2+\alpha}$:
 \begin{equation}
\label{mom1xsubO2palpha}
  \langle x \rangle =
  \langle x_0 \rangle
  -\frac{2  \mu_0  \langle x_0 \rangle  D_\alpha }{(1+\alpha)\Gamma(\alpha)} \, t^{1+\alpha}
  +\frac{4 \mu_0^2  \langle x^3_0 \rangle  D_\alpha }{(2+\alpha)\Gamma(\alpha)}  \, t^{2+\alpha}+o\left( t^{2+\alpha}\right).
\end{equation}
It is easy to see that the insertion of Eqs.~\eqref{mom1xsubOalpha}  and~\eqref{mom3xsubOalpha} into Eq.~\eqref{mom1sub} leads to an equation for $d\langle x\rangle/dt$ to order $t^{1+2\alpha}$, which would in turn lead to an expression for the first-order moment to order  $t^{2+2\alpha}$, and so on.

Let us briefly discuss the difference in behavior between the normal diffusive and the subdiffusive case. From Eq.~\eqref{mom1xsubO2palpha}, we immediately see that the leading correction to the case of a static domain is of the order $t^{1+\alpha}$, i.e., stronger than the quadratic correction predicted by Eq.~\eqref{mom1sO4}. This reflects the fact that the domain growth plays a more dominant role when the intrinsic particle motion is subdiffusive rather than diffusive. The result corresponding to normal diffusion is recovered in the limit $\alpha\to 1$.

The short time behavior of even-order moments can be obtained in a similar way.
 The equations for the first two even-order moments are [c.f. Eqs.~\eqref{mom2} and~\eqref{mom4}]:
\begin{align}
\label{mom2sub}
\frac{d \langle x^2 \rangle}{d t}&=D_\alpha \left[
2 \,_0\mathcal{D}_t^{1-\alpha} \langle x^{0} \rangle
-8 \mu_0 t \,_0\mathcal{D}_t^{1-\alpha} \langle x^{2} \rangle
+12 \mu_0^2 t^2 \,_0\mathcal{D}_t^{1-\alpha} \langle x^{4} \rangle+\cdots
 \right], \\
\label{mom4sub}
\frac{d \langle x^4 \rangle}{d t}&=D_\alpha \left[
12  \,_0\mathcal{D}_t^{1-\alpha} \langle x^{2} \rangle
-32 \mu_0 t \,_0\mathcal{D}_t^{1-\alpha} \langle x^{4} \rangle+\cdots
 \right].
\end{align}
To order $t^{-1+\alpha}$, the equation for $\langle x^2\rangle$  reads
\begin{equation}
\frac{d \langle x^2 \rangle}{d t}=2 D_\alpha
  \,_0\mathcal{D}_t^{1-\alpha} 1  +o\left( t^{-1+\alpha}\right),
\end{equation}
so that
\begin{equation}
\label{mom2xsub1palpha}
 \langle x^2 \rangle =\langle x_0^2\rangle+\frac{2 D_\alpha}{\Gamma(1+\alpha)}\, t^\alpha +o\left( t^{\alpha}\right).
\end{equation}
Note that, to this order, the expansion plays no role.

Assume that all the subdiffusive particles are initially placed at $x=0$.  In this case $\langle x^n_0\rangle=0$ for $n\ge 1$. Inserting this expression into Eq.~\eqref{mom4sub}, one sees that the equation for $d\langle x^4\rangle/dt$ to order $t^{-1+2\alpha}$ takes the form
\begin{equation}
\frac{d \langle x^4 \rangle}{d t}=\frac{24}{\Gamma(2 \alpha)}\,  D_\alpha^2 \, t^{-1+2\alpha}+o\left( t^{-1+2\alpha}\right).
\end{equation}
Consequently,
\begin{equation}
\label{short-time-x4}
 \langle x^4 \rangle =\frac{24 D_\alpha^2}{\Gamma(1+2\alpha)}\,  t^{2\alpha}+o\left( t^{2\alpha}\right).
\end{equation}

An improved differential equation for $\langle x^2 \rangle$ can be obtained by taking advantage of the fact that we now know a more accurate expression for $\langle x^2 \rangle$ [c.f., Eq.~\eqref{mom2xsub1palpha}] which can be inserted into Eq.~\eqref{mom2sub}. The resulting equation is
\begin{equation}
\label{eqsub2alpha}
\frac{d \langle x^2 \rangle}{d t}=\frac{2 D_\alpha}{\Gamma(\alpha)}\,  t^{-1+\alpha}  -\frac{16  D_\alpha^2 \, \mu_0}{\Gamma(2\alpha)}\,    t^{2\alpha}  +o\left( t^{2\alpha}\right).
\end{equation}
Note that, to this order, it is not necessary to include the moment $\langle x^4 \rangle$ (and, a fortiori, higher order moments), since the contribution coming from this term will be at least of order $t^{1+3\alpha}$.   The integration of Eq.~\eqref{eqsub2alpha} yields
\begin{equation}
\label{eqsub2alpha2}
  \langle x^2 \rangle =\frac{2 D_\alpha}{\Gamma(1+\alpha)} \,t^{\alpha}  -\frac{16  D_\alpha^2 \mu_0}{(1+2\alpha)\Gamma(2\alpha)} \,   t^{1+2\alpha}  +o\left( t^{1+2\alpha}\right).
\end{equation}

\begin{figure}
\includegraphics[width=0.49\textwidth]{{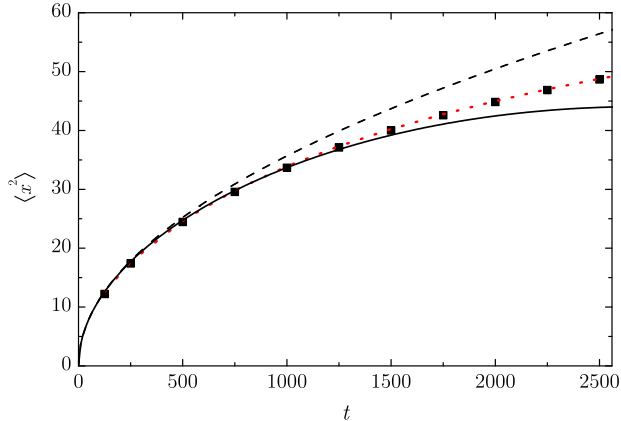}}
\caption{$\langle x^2\rangle$ versus time for $\mu(x,t)=\mu_0x^2, u(x,0)=\delta(x), D=1/2$ and $\mu_0=10^{-6}$ for subdiffusive particles with $\alpha=1/2$. The dotted line represents numerical results, whereas the squares are simulation results. The broken line corresponds to the static case, i.e., to the main term in Eq.~\eqref{eqsub2alpha}, whereas the solid line corresponds to the analytical short time approximation given by the full equation \eqref{eqsub2alpha}.
}
\label{fig10}
\end{figure}

The main term is simply the standard exact expression for the mean square displacement of a subdiffusive particle evolving on a static domain. In Fig.~\ref{fig10} we compare the approximation given by Eq.~\eqref{eqsub2alpha2} with results obtained by numerical integration and by random walk simulations. From Eq.~\eqref{eqsub2alpha2}, one clearly sees that the leading correction introduced by the domain evolution is of the order $t^{1+2\alpha}$, as opposed to the cubic correction characteristic of the normal diffusive case
[cf. Eq.~\eqref{mom2sO3delta0}]. Once again, the limit $\alpha\to 1$ yields the result for normal diffusion.

The short-time behavior of the moments of $u(y,t)$ for subdiffusion is readily obtained by the method already used in the normal diffusive case. Note that Eqs.~\eqref{short-time-y}-\eqref{short-time-y2} continue to hold in the subdiffusive case. Inserting Eqs.~\eqref{mom3xsubOalpha},~\eqref{mom1xsubO2palpha},~\eqref{short-time-x4} and~\eqref{eqsub2alpha2} into Eqs.~\eqref{short-time-y}-\eqref{short-time-y2}, we find
\begin{equation}
\langle y \rangle =
\langle y_0 \rangle
+\frac{1}{3}\mu_0 \langle y_0^3\rangle t
+\frac{2 \mu_0 \langle y_0\rangle}{ \Gamma(2+\alpha)} D_\alpha t^{1+\alpha}
+ \frac{\mu_0^2}{10} \langle y_0^5\rangle \, t^2+ o(t^2)
\end{equation}
for $y_0\neq 0$. Note that the linear drift term already encountered in the normal diffusive case remains dominant.
For $y_0=0$, one has
\begin{equation}
\langle y^2\rangle=
\frac{2D_\alpha}{\Gamma(1+\alpha)} t^\alpha
+ \frac{16 \mu_0 D_\alpha^2}{\Gamma(2+2\alpha)} t^{1+2\alpha}
+o\left( t^{1+2\alpha}\right),
\end{equation}
implying that the first correction arising from the domain evolution is proportional to $t^{1+2\alpha}$, and thus more important than the cubic correction given by Eq.~\eqref{centered-y2}.

\section{Evolution equations for diffusion and reactions on evolving domains}
In the preceding section we derived the evolution equations for the diffusion limits of the master equations for an auxiliary CTRW with standard diffusion, and with subdiffusion.
The auxiliary CTRW was defined as a CTRW on a fixed domain, with variable jump lengths, corresponding to a CTRW on an evolving domain with fixed nearest neighbour jump lengths.
The evolution equations were obtained in the co-ordinate system of the fixed domain.
In this section we obtain the corresponding evolution equations in the co-ordinate system of the evolving domain by mapping the equations, Eq.~(\ref{standard-diff}) for standard diffusion, and Eq.~\eqref{sub} for subdiffusion back to the growing domain. It is important to note that this is largely a formality.
The governing equations are easier to solve for the auxiliary process on the fixed domain.

We recall that the space co-ordinate $y$ on the evolving domain can be represented by $\bar g(x,t)$,
and we have defined functions on the growing domain, $\zeta(y,t)=\zeta(\bar g(x,t),t)=\bar\zeta(x,t)$ in terms of functions on the fixed domain using the mapping, Eq.~\eqref{eq_grow_transform}.
In this way we identify:
\begin{eqnarray}
u(y,t)&=&\bar u(x,t),\\
\mu(y,t)&=&\bar\mu(x,t),\\
\nu(y,t)&=&\bar{\nu}(x,t)=\frac{\partial \bar g(x,t)}{\partial x}=e^{\int_0^t \bar{\mu}(x,s)ds},\\
\eta(y,t)&=&\bar{\eta}(x,t)=\frac{\partial \bar g(x,t)}{\partial t}=\int_0^x \bar{\mu}(z,t) e^{\int_0^t \bar{\mu}(z,s)ds}dz.
\end{eqnarray}
Many results can now be obtained in a straightforward way from simple chain rules, for example,
\begin{equation}
\frac{\partial\bar\mu(x,t)}{\partial x}=\frac{\partial \mu(y,t)}{\partial y}\nu(y,t).
\end{equation}
Note that
\begin{equation}
\frac{\partial \bar \nu(x,t)}{\partial x}=\left(\int_0^t\frac{\partial \bar\mu(x,s)}{\partial x}\, ds\right) e^{\int_0^t \bar\mu(x,s)\, ds}=\left(\int_0^t\frac{\partial \bar\mu(x,s)}{\partial x}\, ds\right)\bar\nu(x,t),
\end{equation}
and
\begin{equation}
\frac{\partial \bar \nu(x,t)}{\partial x}=\left(\frac{\partial \nu(y,t)}{\partial y}\right)\frac{\partial y}{\partial x}=\left(\frac{\partial \nu(y,t)}{\partial y}\right)\bar\nu(x,t),
\end{equation}
so that
\begin{equation}
\int_0^t\frac{\partial \bar\mu(x,s)}{\partial x}\, ds=\frac{\partial \nu(y,t)}{\partial y}.
\end{equation}
We also note that
\begin{equation}
\frac{\partial \bar\nu(x,t)}{\partial t}
=\bar\mu(x,t)e^{\int_0^t \bar\mu(x,s)ds}=\bar\mu(x,t)\bar\nu(x,t),
\end{equation}
and then
\begin{equation}\label{mueta}
\frac{\partial \bar\nu(x,t)}{\partial t}
=\mu(y,t)\nu(y,t).
\end{equation}

Partial derivatives, with respect to time, of functions on the fixed domain  are replaced with material derivatives of the corresponding functions on the growing domain. Thus, if $\zeta(y,t)=\bar\zeta(x,t)$, then
\begin{equation}\label{dzetabardt}
\frac{\partial \bar \zeta(x,t)}{\partial t} =
\frac{\partial \zeta(y,t)}{\partial t}+\frac{\partial \zeta(y,t)}{\partial y}\frac{\partial y}{\partial t}.
\end{equation}
Finally in this preamble we note the following result, obtained by combining Eqs.~(\ref{mueta}) and~(\ref{dzetabardt}),
\begin{equation}
\mu(y,t)=\frac{1}{\nu(y,t)}
\left(\frac{\partial\nu(y,t)}{\partial t}+\frac{\partial \nu(y,t)}{\partial y}\eta(y,t)\right).
\end{equation}

\subsection{Standard Diffusion}
Using the results presented earlier in Section V, it is now straightforward to transform the evolution equation for the auxiliary process with standard diffusion, Eq.~\eqref{standard}, to an evolution equation in the co-ordinates of the evolving domain. There are various ways to express the resulting equation,
\begin{eqnarray}\label{eq_special1}
\frac{\partial u(y,t)}{\partial t}&=&D\frac{\partial^2 u(y,t)}{\partial y^2}-\eta(y,t)\frac{\partial u(y,t)}{\partial y}\nonumber\\
& &-(\mu(y,t)+\omega(y,t))u(y,t)
+\chi(y,t),
\end{eqnarray}
or
\begin{eqnarray}
\label{eq_special2}
\frac{\partial u(y,t)}{\partial t}&=&D\frac{\partial^2 u(y,t)}{\partial y^2}-\eta(y,t)\frac{\partial u(y,t)}{\partial y}\nonumber\\
& &+\left(\frac{1}{\sigma(y,t)}\frac{\partial \sigma(y,t)}{\partial t}+\frac{\partial \sigma(y,t)}{\partial y}\eta(y,t)\right)u(y,t)\nonumber\\
& &
+\chi(y,t),
\end{eqnarray}
or
\begin{eqnarray}\label{eq_special3}
\frac{\partial u(y,t)}{\partial t}&=&D\frac{\partial^2 u(y,t)}{\partial y^2}-\eta(y,t)\frac{\partial u(y,t)}{\partial y}\\
& &-\left(\frac{1}{\nu(y,t)}\frac{\partial \nu(y,t)}{\partial t}+\frac{1}{\nu(y,t)}\frac{\partial \nu(y,t)}{\partial y}\eta(y,t)\right)u(y,t)\nonumber\\
& &-\omega(y,t)u(y,t)
+\chi(y,t).
\end{eqnarray}

\subsection{Subdiffusion}
We now carry out the  transformation of the evolution equation for the auxiliary process with standard diffusion, Eq.~\eqref{sub}, to an evolution equation in the co-ordinates of the evolving domain. In this transformation, the fractional derivatives with respect to time of functions on the fixed domain are replaced with co-moving fractional derivatives on the evolving domain, defined as \cite{AHM2017pre},
 \begin{equation}
 \label{eq_new_fd}
\,_0^{g}\mathcal{C}_t^{1-\alpha}f(y,t)=\frac{1}{\Gamma(\alpha)}\frac{\partial}{\partial t}\int_0^t f(\bar g(\bar g^{-1}(y,t),t'),t')(t-t')^{\alpha-1}dt'.
 \end{equation}
The evolution equation for subdiffusion with reactions on the evolving domain can now be written as
\begin{equation}
\label{eq_final}
\begin{split}
\frac{\partial u(y,t)}{\partial t}=&D_\alpha\frac{\partial^2}{\partial y^2}\left(\sigma(y,t,0)\,_0^{g}\mathcal{C}_t^{1-\alpha}\left(\frac{u(y,t)}{\sigma(y,t,0)}\right)\right)-\eta(y,t)\frac{\partial u(y,t)}{\partial y}\\
&-(\mu(y,t)+\omega(y,t))u(y,t)+\chi(y,t).
\end{split}
\end{equation}

Note that if we take the limit as $\alpha \rightarrow 1$ in Eq.~\eqref{eq_final}, we recover the equation for standard diffusion with reactions on an evolving domain, Eq.~\eqref{eq_special1}. Note also that if the growth rate is zero, $\bar\mu(x,t)=0$, then $\eta(y,t)=0$ and $y=x$, and Eq.~\eqref{eq_final} reduces to the equation for subdiffusive
transport with reactions on a fixed domain \cite{SSS2006,LHW2008}, i.e.,
\begin{equation}\label{special4}
\begin{split}
\frac{\partial u(x,t)}{\partial t}=&D_\alpha\frac{\partial^2}{\partial x^2}\left(\theta(x,t,0)\,_0\mathcal{D}_t^{1-\alpha}\left(\frac{u(x,t)}{\theta(x,t,0)}\right)\right)\\
&-\omega(x,t)u(x,t)+\chi(x,t).
\end{split}
\end{equation}

Finally we note that if there are no reactions then Eq.~\eqref{eq_final} reduces to
\begin{equation}
\label{eq_final2}
\begin{split}
\frac{\partial u(y,t)}{\partial t}=&D_\alpha\frac{\partial^2}{\partial y^2}\left(e^{-\int_0^t\bar\mu(\bar g^{-1}(y,t),s)\, ds}\,_0^{g}\mathcal{C}_t^{1-\alpha}\left(\frac{u(y,t)}{e^{-\int_0^t\bar\mu(\bar g^{-1}(y,t),s)\, ds}}\right)\right)-\eta(y,t)\frac{\partial u(y,t)}{\partial y}\\
&-\mu(y,t)u(y,t).
\end{split}
\end{equation}
This is in agreement with Eq.~(40) in Ref.~\cite{AHM2017pre} in the special case of uniform growth where $\mu(y,t)=r$ and $y(x,t)=xe^{rt}$ so that $\eta(y)=ry$.

\section{Example}

In  this section, we will show how our formalism could be used for describing a specific reaction-diffusion process occurring in a growing surface. In order to apply the equations in this paper, it is necessary to have explicit expressions for the domain growth function $\bar\mu(x,t)$ and the birth and death rates $\bar\chi(x,t)$ and $\bar\omega(x,t)$.
As an illustration we consider a population with logistic growth dynamics, spreading along an evolving interface whose height $h(r,t)$ above a horizontal baseline $r\in[0,L]$ is given by
\begin{equation}
\label{hr}
\frac{\partial h(r,t)}{\partial t}=v+\lambda\left(1+\frac{1}{2}\left(\frac{\partial h(r,t)}{\partial r}\right)^2\right).
\end{equation}
In this model, $v$ represents a vertical growth rate parameter, $\lambda$ represents a surface normal growth rate parameter, and it is assumed that the height is a slowing varying function of $r$, i.e., $\partial h/\partial r\ll 1$.
This surface growth model has been used to model laminations in growing stromatolites \cite{BBHW2000,BBHJ2004}.
The application considered here could model the lateral spread of microbicides on the surface of a growing stromatolite.

The arc length along the interface is given by
\begin{equation}
s(r,t)=\int_0^r\sqrt{1+\left(\frac{\partial h(r',t)}{\partial r'}\right)^2}\, dr'.
\end{equation}
The initial fixed domain co-ordinate $x\in [0,L_0]$ can then be defined as
\begin{equation}\label{xs}
x=s(r,0),
\end{equation}
and the evolving domain co-ordinate $y\in [0,L(t)]$ is given as
\begin{equation}\label{ys}
y(x,t)=s(r(x),t).
\end{equation}
Note that Eq.~\eqref{xs} defines $r(x)$ used in Eq.~\eqref{ys}.
The growth rate $\bar\mu(x,t)$ can then be obtained by differentiating Eq.~\eqref{ymux} with respect to $x$, taking the logarithm, and differentiating with respect to $t$,
\begin{equation}
\label{sbarmu}
\bar\mu(x,t)=\frac{\partial}{\partial t}\log\left(\frac{\partial y}{\partial x}\right).
\end{equation}

In logistic growth a population with number density $u(x,t)$ evolves via
\begin{equation}
\frac{\partial u(x,t)}{\partial t}=\gamma u(x,t)\left(1-\frac{u(x,t)}{u_0}\right),
\end{equation}
where $u_0$ is the threshold carrying capacity and $\gamma$ is a net per capita growth rate.
This then identifies
\begin{equation}
\label{exchiomega}
 \bar\chi(x,t)=\gamma u(x,t)\quad\mbox{and}\quad \bar\omega(x,t)=\frac{\gamma  u(x,t)}{u_0}.
 \end{equation}

The expressions for $\bar\mu(x,t)$, $\bar\omega(x,t)$ and $\bar\chi(x,t)$ can be employed in the evolution equations of $u(x,t)$ with standard diffusion, Eq.~\eqref{u-standard-diff}, or subdiffusion, Eq.~\eqref{sub3}, and the solutions $u(x,t)$  can be mapped onto corresponding locations $y$ at time $t$ on the evolving domain using Eq.~\eqref{ys}.
The solutions of Eq.~\eqref{u-standard-diff} or Eq~\eqref{sub3} could be obtained using numerical methods such as a modification of the fractional Crank-Nicolson of Ref.~\cite{Yuste2006} or a modification of the discrete time random walk algorithm presented in Ref.~\cite{angstmann2016stochastic}.

\section{Summary and Outlook}
In this paper we have derived evolution equations for a system undergoing diffusion and reactions on an arbitrarily evolving one-dimensional domain. Evolution equations have been obtained for both standard diffusion and subdiffusion.
The evolution equations were obtained from the following sequence of steps. First we identified a general mapping between the initial fixed domain and the evolving domain. We then derived master equations for an auxiliary CTRW on a fixed domain with birth and death processes, corresponding to a CTRW with unbiased fixed length steps on a growing domain with birth and death processes. We considered particular waiting time densities, corresponding to standard diffusion in one case and subdiffusion in another,  and we obtained diffusion limits of the master equations for the auxiliary process on the fixed domain. We then mapped the governing equations back to the evolving domain.

We developed an iterative method for obtaining analytic expressions for short time moments for standard diffusion and for subdiffusion on an evolving domain. We showed that the moments calculated in this way compared favourably with moment evaluations obtained from numerical solutions of the governing equations, and from numerical simulations of the processes.

Beyond this quantitative agreement, let us briefly enumerate some key features of the underlying physics. We have seen that the short-time behavior of the moments in the absence of reactions is characterized by the strong influence of the initial condition. In this paper, we have largely focused on particular case $\mu=\mu_0 x^2$ (with $\mu_0>0$, say) which lends itself particularly well to analytical treatment. In terms of $x$-coordinates, when a particle starts away from $x=0$, it is dragged towards the origin in an accelerated fashion. In the case of normal diffusion, the behavior of $\langle x^2\rangle$ is characterized by a correction to the standard contribution $2Dt$ which is quadratic in time and which also contains the diffusion coefficient $D$. When the particle starts at the origin, the correction to the diffusive contribution is quadratic rather than cubic.  For any starting point $y_0\neq 0$, the behavior of $\langle y \rangle$ is characterized by a linear correction which does not depend on the diffusivity $D$.  The apparent diffusivity in $y$-coordinates is also modified by an additive contribution that stems from the domain evolution. In the subdiffusive case, the corrections associated with the domain evolution are stronger, since the relative influence of subdiffusive transport is less important than that of Brownian diffusion.

As an application  we considered how a diffusion process in an evolving domain corresponding to surface growth, above a horizontal baseline, driven by constant vertical growth and surface normal growth could be described in terms of the formalism developed in this paper. The co-ordinates for the evolving domain represent the arc length along the surface in this application. A general framework for applications would require extensions in higher spatial dimensions, consideration of different boundary conditions, inclusion of forces, and inclusion of stochastic fluctuations.

We close by noting that the obtained results may be relevant for the study of encounter-controlled reactions in non-uniformly evolving domains. Consider, for example, two pulses describing two diffusing particles that evolve on a domain with expansion rate $\mu\propto x^2$.  Further, assume that these particles react instantaneously upon encounter. Strictly speaking, the computation of the reaction rate is a first-passage problem, but an approximation based on the overlap of the two pulses (defined via the respective pulse widths/second order moments) may yield acceptable results for a suitable parameter choice.  If the initial separation is small enough, the reaction rate will be dominated by the short-time regime, and some of our results might prove useful. As we have seen, for a given initial separation, the reaction rate may display significant differences depending on the initial location of the pulses with respect to the origin.  More generally, the study of first passage problems in non-uniformly evolving domains may be of interest for biological applications. An example are target and trapping problems, such as the subdiffusive trapping problem studied in
Ref.~\cite{YusteAcedo2004}. An extension of the results available in the literature for the case of a static domain may unveil interesting effects arising from the interplay between intrinsic transport and domain evolution.

\section*{Acknowledgments}

C.~N.~A., and B.~I.~H. acknowledge support by the Australian Commonwealth Government (ARC DP200100345). E.~A., F.~L.~V., and S.~B.~Y.   acknowledge support by the Spanish  Agencia Estatal de Investigaci\'on Grant (partially financed by the ERDF) No.~FIS2016-76359-P and by the Junta de Extremadura (Spain) Grant (also partially financed by the ERDF) No.~GR18079. In addition, F.~L.~V. acknowledges financial support from the Junta de Extremadura through Grant No. PD16010 (partially financed by FSE funds).

\appendix*

\section{Non-uniform nearest neighbour steps on a fixed domain}

Here we relate nearest neighbour jump lengths of a fixed size $\Delta y$ on the growing domain to corresponding jump lengths $\epsilon^+$ and $\epsilon^-$ for the auxiliary CTRW on the original fixed domain. The jump lengths on the $\epsilon^+$ and $\epsilon^-$ were defined in Sec.~\ref{secDifLimAuxCTRW}  by the relations $y-\Delta y=\bar g(x-\epsilon^-,t)$ and  $y+\Delta y=\bar g(x+\epsilon^+,t)$. Taking into account that $y=\bar g(x,t)$, we can rewrite these expressions as
\begin{equation}
\pm \Delta y=  \bar g(x\pm\epsilon^\pm,t)- \bar g(x,t).
\end{equation}
We now take Taylor series expansions around the point $x$ and retain leading order terms in $\epsilon^+$ and $\epsilon^-$ to arrive at
\begin{equation}
\Delta y=\epsilon^\pm \bar g(x,t)\pm\frac{(\epsilon^\pm)^2}{2}\bar g_{xx}(x,t)+O((\epsilon^\pm)^3).
\end{equation}
We can now solve the above quadratic approximations for $\epsilon^+$ and $\epsilon^-$, noting that both terms must vanish when $\Delta y=0$, to arrive at
\begin{equation}
\epsilon^\pm \approx \frac{\mp \bar g_x(x,t)\pm \bar g_x(x,t)\sqrt{1+\frac{2\bar g_{xx}(x,t)}{\bar g_x(x,t)^2}\Delta y}}{\bar g_{xx}}.
\end{equation}
We now expand each of these terms as a series expansion in powers of $\Delta y$, arriving at
\begin{equation}
\epsilon^\pm=\frac{\Delta y}{\bar g_x(x,t)}\mp\frac{\bar g_{xx}(x,t)}{2\bar g_x(x,t)^3}\Delta y^2+O(\Delta y^3)
\end{equation}
Finally, using Eq.~\eqref{ymux} we obtain Eqs.~\eqref{epspm}.


\begin{thebibliography}{10}

\bibitem{O1980}
A.~Okubo, Diffusion and ecological problems: Mathematical models,
  Biomathematics \textbf{10}, 114 (1980).

\bibitem{B1986}
N.~F. Britton, \emph{Reaction-diffusion equations and their applications to
  biology} (Academic Press, London, 1986).

\bibitem{M2001}
J.~D. Murray, \emph{Mathematical Biology. II Spatial Models and Biomedical
  Applications} (Springer, New York, 2003).

\bibitem{HW2000}
B.~I. Henry and S.~L. Wearne, Fractional reaction-diffusion, Physica A \textbf{276}, 448 (2000).

\bibitem{HLW2006}
B.~I. Henry, T.~A.~M. Langlands, and S.~L. Wearne, Anomalous diffusion with linear
reaction dynamics: From continuous time random walks to fractional
reaction-diffusion equations, Phys. Rev. E \textbf{74}, 031116 (2006).

\bibitem{SSS2006}
I.~M. Sokolov, M.~G.~W. Schmidt, and F.~Sagu{\'e}s, Reaction-subdiffusion
  equations, Phys. Rev. E \textbf{73}, 031102 (2006).

\bibitem{LHW2008}
T.~A.~M. Langlands, B.~I. Henry, and S.~L. Wearne, Anomalous subdiffusion with
  multispecies linear reaction dynamics, Phys. Rev. E \textbf{77}, 021111 (2008).

\bibitem{F2010}
S.~Fedotov, Non-Markovian random walks and nonlinear reactions: Subdiffusion
  and propagating fronts, Phys. Rev. E \textbf{81},  011117 (2010).

\bibitem{AYL2010}
E.~Abad, S.~B. Yuste, and K.~Lindenberg, Reaction-subdiffusion and
  reaction-superdiffusion equations for evanescent particles performing
  continuous-time random walks, Phys. Rev. E \textbf{81}, 031115 (2010).

\bibitem{YAL2010}
S.~B. Yuste, E.~Abad, and K.~Lindenberg, Reaction-subdiffusion model of morphogen
  gradient formation, Phys. Rev. E \textbf{82}, 061123 (2010).

\bibitem{ADH2013mmnp}
C.~N. Angstmann, I.~C. Donnelly, and B.~I. Henry, Continuous time random walks with
  reactions forcing and trapping, Math. Model. Nath. Phenom. \textbf{8}, 17 (2013).

\bibitem{CGK1999}
E.~J. Crampin, E.~A. Gaffney, and P.~K. Maini, Reaction and diffusion on growing
  domains: scenarios for robust pattern formation, Bull. Math. Biol. \textbf{61},
  1093 (1999).

\bibitem{CM2001}
E.~J. Crampin and P.~K. Maini, Modelling biological pattern formation: the role of
  domain growth, Comments Theor. Biol. \textbf{6}, 229 (2001).

\bibitem{BYE2010}
R.~E. Baker, C.~A. Yates, and R.~Erban, From microscopic to macroscopic
  descriptions of cell migration on growing domains, Bull. Math. Biol. \textbf{72},
  719 (2010).

\bibitem{WBGM2011}
T.~E. Woolley, R.~E. Baker, E.~A. Gaffney, and P.~K. Maini, Stochastic reaction and
  diffusion on growing domains: understanding the breakdown of robust pattern
  formation, Phys. Rev. E \textbf{84},  046216 (2011).

\bibitem{YBEM2012}
C.~A. Yates, R.~E. Baker, R.~Erban, and P.~K. Maini, Going from microscopic to
  macroscopic on nonuniform growing domains, Phys. Rev. E \textbf{86}, 021921 (2012).

\bibitem{SSMB2015}
M.~J. Simpson, J.~A. Sharp, L.~C. Morrow, and R.~E. Baker, Exact solutions of
  coupled multispecies linear reaction-diffusion equations on a uniformly
  growing domain, PLoS One \textbf{10},  e0138894 (2015).

\bibitem{YAE2016}
S.~B. Yuste, E.~Abad, and C.~Escudero, Diffusion in an expanding medium:
Fokker-Planck equation, Green's function, and first-passage properties,
Phys. Rev. E \textbf{94}, 032118 (2016).

\bibitem{MB2018}
A.~Madzvamuse and R.~Barreira,
Domain-growth-induced patterning for reaction-diffusion systems with linear cross-diffusion,
Discrete Cont. Dyn. Syst. B \textbf{23}, 2775 (2018).

\bibitem{VEAY2018}
F.~Le Vot, C.~Escudero, E.~ Abad, and S.~B.~Yuste,
Encounter-controlled coalescence and annihilation on a one-dimensional growing domain,
Phys. Rev. E \textbf{98} 032137 (2018).

\bibitem{GKK2019}
R.~A.~Van Gorder, V.~Klika, and A.~L.~Krause,
Turing conditions for pattern forming systems on evolving manifolds,
 arXiv:1904.09683v2 [nlin.PS].

\bibitem{BWM2000}
D.~A. Benson, S.~W. Wheatcraft, and M.~M. Meerschaert, The fractional-order
  governing equation of L{\'e}vy motion, Water Resour. Res. \textbf{36}, 1413 (2000).

\bibitem{LB2003}
M.~Levy and B.~Berkowitz, Measurement and analysis of non-fickian dispersion in
  heterogeneous porous media, J. Contam. Hydrol. \textbf{64}, 203 (2003).

\bibitem{WEKN2004}
M.~Weiss, M.~Elsner, F.~Kartberg, and T.~Nilsson, Anomalous subdiffusion is a
  measure for cytoplasmic crowding in living cells, Biophys. J. \textbf{87}, 3518 (2004) .

\bibitem{SWDA2006}
F.~Santamaria, S.~Wils, E.~{De Schutter}, G.~J. Augustine, Anomalous diffusion
  in Purkinje cell dendrites caused by spines, Neuron \textbf{52},  635 (2006) .

\bibitem{MW2010}
N.~Malchus and M.~Weiss, Elucidating anomalous protein diffusion in living cells
  with fluorescence correlation spectroscopy: facts and pitfalls, J. Fluoresc. \textbf{20},  19 (2010).

\bibitem{AYYHY2011}
T.~Akimoto, E.~Yamamoto, K.~Yasuoka, Y.~Hirano, and M.~Yasui, Non-Gaussian
  fluctuations resulting from power-law trapping in a lipid bilayer, Phys.
  Rev. Lett. 107 178103 (2011).

\bibitem{EK2011}
I.~Eliazar and J.~Klafter, Anomalous is ubiquitous, Ann. Phys. \textbf{326}, 2517 (2011).

\bibitem{S2012}
I.~M. Sokolov, Models of anomalous diffusion in crowded environments, Soft
  Matter \textbf{8}, 9043 (2012) .

\bibitem{JTHDB2016}
R.~P. Joyner, J.~H. Tang, J.~Helenius, E.~Dultz, C.~Brune, L.~J. Holt, S.~Huet,
  D.~J. Mueller, and K.~Weis, A glucose-starvation response regulates the diffusion
  of macromolecules, Elife \textbf{5}, e09376 (2016).

\bibitem{VAY2017}
F.~L. Vot, E.~Abad, and S.~B. Yuste, Continuous time random walk model for anomalous
  diffusion in expanding media, Phys. Rev. E \textbf{96}, 032117 (2017).

\bibitem{VY2018}
F.~L. Vot, S.~B. Yuste, Continuous time random walks and Fokker-Planck equation
  in expanding media, Phys. Rev. E \textbf{98}, 042117 (2018).

\bibitem{AHM2017pre}
C.~N. Angstmann, B.~I. Henry, and A.~V. McGann, Generalized fractional diffusion
  equations for subdiffusion in arbitrarily growing domains, Phys. Rev. E \textbf{96}, 042153 (2017).

\bibitem{MW1965}
E.~Montroll, G.~Weiss, Random walks on lattices II, J. Math. Phys. \textbf{6}, 167 (1965).

\bibitem{MK2000}
R.~Metzler, J.~Klafter, The random walk's guide to anomalous diffusion: A
  fractional dynamics approach, Phys. Rep. \textbf{339}, 1 (2000).

\bibitem{EYAV2018}
C.~Escudero, S.~B. Yuste, E.~Abad, F.~L. Vot, Reaction-diffusion kinetics in
  growing domains, Handbook of Statistics \textbf{39}, 131 (2018).

\bibitem{AEVY2019}
E.~Abad, C.~Escudero,  F.~L. Vot, and S.~B. Yuste, First-Passage Processes and Encounter-Controlled Reactions in Growing Domains, in \emph{Chemical Kinetics: Beyond the Textbook}, edited by K. Lindenberg, R. Metzler, and G. Oshanin, pp. 409-433 (World Scientific, Singapore, 2019).

\bibitem{Podlubny1999}  I. Podlubny, \emph{Fractional Differential Equations: An Introduction to Fractional Derivatives, Fractional Differential Equations, to Methods of Their Solution and Some of Their Applications} (Academic Press, San Diego, 1999).

\bibitem{LVYA2019} F.~L. Vot, E.~Abad, and S.~B. Yuste, Standard and fractional Ornstein-Uhlenbeck process on a growing domain, Phys. Rev. E \textbf{100}, 012142 (2019).

\bibitem{Yuste2006} S.~B.~Yuste, Weighted average finite difference methods for fractional diffusion equations,
J. Comp. Phys. \textbf{216}, 264 (2006).

\bibitem{BBHW2000}
M.~Bachelor, R.~Burne, B.~Henry, and S.~Watt, Deterministic KPZ model for
  stromatolite laminae, Physica A \textbf{282}, 123 (2000).

\bibitem{BBHJ2004}
M.~Bachelor, R.~Burne, B.~Henry, and M.~Jackson, A case for biotic morphogenesis of
  coniform stromatolites, Physica A \textbf{337}, 319 (2004).

\bibitem{angstmann2016stochastic}
C.~N. Angstmann, I.~C. Donnelly, B.~I. Henry, B.~A. Jacobs, T.~A.~M. Langlands,
and J.~A. Nichols, From stochastic processes to numerical methods: A new scheme
for solving reaction subdiffusion fractional partial differential equations,
J. Comput. Phys. \textbf{307}, 508 (2016).

 \bibitem{YusteAcedo2004}
S.~B. Yuste and L. Acedo, Some exact results for the trapping of subdiffusive particles in one dimension,
Physica A \textbf{336}, 334 (2004).

\end{thebibliography}
\end{document}